\documentclass[11pt]{article}
\usepackage{amsmath,amssymb,amsfonts,amsthm,enumerate,esint}
\usepackage{moreverb,rotating,graphics}
\usepackage[T1]{fontenc}
\usepackage[utf8]{inputenc}
\usepackage[francais,english]{babel} 
\usepackage{dsfont}
\usepackage{float}
\usepackage{mdframed}
\usepackage{subfig}
\usepackage{color}
\usepackage{multirow}
\usepackage{hyperref} 
\usepackage{authblk}
\usepackage[a4paper]{geometry}
\hypersetup{colorlinks=true, linkcolor =red, citecolor = blue}

\newtheorem{algorithm}{Algorithm}

\def\tr{{\rm Tr \,}}

\def\N{{\mathbb N}}
\def\Z{{\mathbb Z}}

\def\R{{\mathbb R}}
\def\C{{\mathbb C}}

\def\cD{{\mathcal D}}

\def\cH{{\mathcal H}}
\def\cI{{\mathcal I}}
\def\cJ{{\mathcal J}}
\def\cK{{\mathcal K}}

\def\cM{{\mathcal M}}

\def\cP{{\mathcal P}}
\def\cQ{{\mathcal Q}}
\def\cR{{\mathcal R}}

\def\cV{{\mathcal V}}

\def\balpha{{\boldsymbol\alpha}}

\def\ba{{\mathbf a}}

\def\bk{{\mathbf k}}
\def\bK{{\mathbf K}}

\def\bq{{\mathbf q}}
\def\br{{\mathbf r}}

\def\bR{{\mathbf R}}

\def\1{{\mathds{1}}}
\newcommand \dps{\displaystyle }

\title{Compression of Wannier functions into Gaussian-type orbitals}
\author[1]{Athmane Bakhta}
\author[2]{Eric Canc\`es}
\author[3]{Paul Cazeaux}
\author[4]{Shiang Fang}
\author[5]{Efthimios Kaxiras}
\affil[1]{Universit\'e Paris Est, CERMICS (ENPC), F-77455, Marne-la-Vall\'ee, France}
\affil[2]{Universit\'e Paris Est, CERMICS (ENPC), INRIA, F-77455, Marne-la-Vall\'ee, France}
\affil[3]{Department of Mathematics, University of Kansas, Lawrence, Kansas 66045-7594, USA}
\affil[4,5]{Department of Physics, Harvard University, Cambridge, Massachusetts 02138, USA}
\date{}                    
\begin{document}
\selectlanguage{english}

\maketitle
\begin{abstract}
We propose a greedy algorithm for the compression of Wannier functions into Gaussian-polynomials orbitals. The so-obtained compressed Wannier functions can be stored in a very compact form, and can be used to efficiently parameterize effective tight-binding Hamiltonians for multilayer 2D materials for instance. The compression method preserves the symmetries (if any) of the original Wannier function. We provide algorithmic details, and illustrate the performance of our implementation on several examples, including graphene, hexagonal boron-nitride, single-layer FeSe, and bulk silicon in the diamond cubic structure.
\end{abstract}

\section{Introduction}
Since their introduction in 1937 \cite{W37}, Wannier functions have become a widely used computational tool in solid state physics and materials science. Theses functions provide insights on chemical bonding in crystalline material \cite{MMYSV12}, they play an essential role in the modern theory of polarization~\cite{king1993theory}, they can be used to parametrize tight-binding Hamiltonians for the calculation of electronic properties~\cite{fang2015abinito}, and are useful in several other applications~\cite{MMYSV12}.

\medskip

Maximally localized Wannier functions (MLWFs) were introduced by Marzari and Vanderbilt \cite{MV97} and are obtained by minimizing some spread functional \cite{MV97,SMV01,MMYSV12}. Several algorithms for generating MLWFs are implemented in the Wannier90 computer program~\cite{W90_code}. In the general case, MLWFs obtained by the standard Marzari-Vanderbilt procedure are not centered at high-symmetry points of the crystal (typically atoms or centers of chemical bonds), and do not fulfill any symmetry properties~\cite{SMV01,THJ05}, which complicates their physical interpretation. Symmetry-adapted Wannier functions (SAWFs) are centered at high-symmetry points and are associated with irreducible representations of a non-trivial subgroup of the space group of the crystal (precise definitions are given in Appendix). They are the solid-state counterparts of symmetry-adapted molecular orbitals~\cite{L16} that are fruitfully used in quantum chemistry. SAWFs were investigated in \cite{Cloizeaux,K73,BC79,K87,SB94,ES97,SU01,SE05,PBMM02,CZP06} from both the theoretical and the numerical point of view. An algorithm for generating maximally-localized SAWFs was recently proposed by Sakuma \cite{S13}, which makes it possible to enforce the center and symmetries of the Wannier functions during the spread minimization procedure, and has been implemented in the Wannier90 package. 

\medskip

 In this work, we propose a numerical method for compressing Wannier functions into a finite sum of Gaussian-polynomial functions, referred to as Gaussian-type orbitals (GTOs), which preserves the centers and the possible symmetries of the original Wannier functions. Such compressed representations enable the characterization of a Wannier function by a small number of parameters (the shape parameters of the Gaussians and the polynomial coefficients) rather than by its values on a potentially very large grid.  In addition, they can be used to accelerate the parameterization of tight-binding Hamiltonians or more advanced reduced models from Wannier functions computed from Density Functional Theory. Indeed, matrix elements of effective Hamiltonians can be computed very efficiently using GTOs; this fundamental remark by Boys~\cite{B50} was instrumental for the development of numerical methods for quantum chemistry. Gaussian-type approximate Wannier functions should be particularly useful for simulating multilayer two-dimensional materials~\cite{JM13,FK16}, especially when Fock exchange terms are considered, which is the case for hybrid functionals.
 
\medskip
This article is organized as follows. In Section~\ref{sec:theory}, we describe our approach for compressing a given symmetry-adapted Wannier function $W$ into a finite sum of GTOs $\widetilde W_p$ sharing the same center and symmetries as $W$. Note that our procedure is also valid if the Wannier function has no symmetry (in this case the symmetry group is reduced to the identity matrix). The main idea is to construct a sequence $\widetilde W_0, \widetilde W_1, \widetilde W_2, \cdots$ of successively better approximations of $W$ (for the relevant metric, see Section~\ref{sec:Sobolev}), by means of an orthogonal greedy algorithm~\cite{temlyakov_rev,cv_greedy}. The basics of greedy algorithms and symmetry-adapted Wannier functions are briefly summarized in Sections~\ref{sec:greedy} and~\ref{sec:SAWF} respectively. An overall description of our algorithm is given in~Section~\ref{sec:alg_greedy} and implementation details are provided in Section~\ref{sec:alg_det}. Greedy methods are very well adapted to the compressing problem under consideration, but our implementation is not necessarily optimal: many variants of the numerical scheme described in Section~\ref{sec:alg_det} can be considered, and there is clearly room for improvement to reduce the number of GTOs necessary to reach a given accuracy. The purpose of this work is to assess the efficiency of greedy methods in this setting, and to  stimulate further work. The performance of our current implementation is illustrated in Section~\ref{sec:num} on four examples: three two-dimensional materials, namely graphene, hexagonal boron-nitride (hBN), and FeSe, and bulk silicon in the cubic diamond structure.

%%%%%%%%%%%%%%%%%%%%%%%%
%%%%%%%%%%%%%%%%%%%%%%%%
\section{Theory}
\label{sec:theory}
%%%%%%%%%%%%%%%%%%%%%%%%
\subsection{Error control}
\label{sec:Sobolev}
Consider a real-valued Wannier function $W : \R^3 \to \R$, which we would like to approximate by a finite sum of well-chosen Gaussian-polynomial functions. First, we have to specify the norm with which the error between $W$ and its approximation $\widetilde W$ will be measured. We will consider here the $L^2$ and $H^1$ norms respectively defined by
$$
\|u\|_{L^2} = \left( \int_{\R^3} |u(\br)|^2 \, d\br \right)^{1/2}
$$
and 
\begin{equation}\label{eq:defH1}
\|u\|_{H^1} = \left( \int_{\R^3} |u(\br)|^2 \, d\br  + \int_{\R^3} |\nabla u(\br)|^2 \, d\br  \right)^{1/2}.
\end{equation}
Requiring that $\|W-\widetilde W\|_{H^1}$ is small is far more demanding than simply requesting that $\|W-\widetilde W\|_{L^2}$ is small. In using approximate Wannier functions to calibrate tight-binding models, it is important to require $\|W-\widetilde W\|_{H^1}$ to be small. Indeed, while the errors on the overlap integrals can be controlled by $L^2$-norms:
$$
\left| \int_{\R^3} W_i(\br)W_j(\br) \, d\br - \int_{\R^3} \widetilde W_i(\br)\widetilde W_j(\br) \, d\br \right| \le \| W_i \|_{L^2} \|W_i-\widetilde W_i\|_{L^2} + \| \widetilde W_i \|_{L^2} \|W_j-\widetilde W_j\|_{L^2},
$$
the errors on the kinetic energy integrals appearing in effective one-body Hamiltonians matrix elements 
$$
\langle W_i | H | W_j \rangle =  \frac 12  \int_{\R^3} \nabla W_i(\br) \cdot \nabla W_j(\br) \; d\br  + \int_{\R^3} \cV(\br) W_i(\br)W_j(\br) \, d\br 
$$
are controlled by the $L^2$-norms of the gradients, hence by the $H^1$-norms of the functions. The $H^1$-norm also allows one to control the errors on the potential integrals, even in presence of Coulomb singularities. Our greedy algorithm has been implemented in the Fourier representation, and can therefore minimize the error between the Wannier function $W$ and its GTO representation for any value of the Sobolev exponent $s$.

\medskip
Note that the $L^2$ and $H^1$-norms are particular instances of the Sobolev norms $H^s$, $s \in \R$, defined on the Solobev spaces
$$
H^s(\R^3) = \left\{ u : \R^3 \to \R \mbox{ s.t. } \int_{\R^3} (1+|\bk|^2)^{s} |\widehat u(\bk)|^2 \, d\bk < \infty \right\},
$$
where $\widehat u$ is the Fourier transform of $u$, by
\begin{equation}\label{eq:defHs}
\| u \|_{H^s}:= \left( \int_{\R^3} (1+|\bk|^2)^{s} |\widehat u(\bk)|^2 \, d\bk \right)^{1/2}.
\end{equation}
The $L^2$-norm corresponds to $s=0$, due to the isometry property of the Fourier transform:
$$
\int_{\R^3}  |\widehat u(\bk)|^2 \, d\bk = \int_{\R^3}  |u(\br)|^2 \, d\br.
$$
Likewise, definition \eqref{eq:defHs} agrees with definition\eqref{eq:defH1} for $s=1$ since 
$$
\int_{\R^3}  |\bk|^2 |\widehat u(\bk)|^2 \, d\bk = \int_{\R^3}   |i \bk\widehat u(\bk)|^2 \, d\bk = \int_{\R^3}   \left| \widehat{\nabla u}(\bk)\right|^2 \, d\bk =   \int_{\R^3} |\nabla u(\br)|^2 \, d\br .
$$ In the numerical examples reported in Section~\ref{sec:num}, we will consider the cases $s=0$ and $s=1$.

%%%%%%%%%%%%%%%%%%%%%%%%
\subsection{Greedy algorithms in a nutshell}
\label{sec:greedy}
Greedy algorithms~\cite{temlyakov_rev,cv_greedy} are iterative algorithms that, among other things, construct sequences of approximations $\widetilde W_0$, $\widetilde W_1$, $\widetilde W_2$, ... of some target function $W \in H^s(\R^3)$, with the following properties:
\begin{itemize}
\item each approximate function $\widetilde W_p$ is a sum of $p$ "simple" functions belonging to some prescribed {\em dictionary} $\cD \subset H^s(\R^3)$:
$$
\widetilde W_p(\br) = \sum_{j=1}^p \phi_j^{(p)}(\br), 
$$
with $\phi_j^{(p)} \in \cD$. In our case, $\cD$ will be a set of symmetry-adapted Gaussian-polynomial functions;
\item the errors $\|W - \widetilde W_p\|_{H^s}$ decay to $0$ when $p \to \infty$.
\end{itemize}
Greedy algorithms therefore provide systematic ways to approximate a given function $W \in H^s(\R^3)$ by a finite sum of simple functions with an arbitrary accuracy. The set $\cD$ of elementary functions cannot be any subset $H^s(\R^3)$ (for instance $\cD$ cannot be chosen as the set of radial functions since only radial functions can be well approximated by finite sums of radial functions). The convergence property $\|W-\widetilde W_p\|_{H^s} \to 0$ is guaranteed provided the set $\cD$ is a {\em dictionary} of $H^s(\R^3)$, that is, a family of functions $H^s(\R^3)$ satisfying the following three conditions:
\begin{enumerate}
\item $\cD$ is a cone, that is, if $\phi \in \cD$, then $t\phi \in \cD$ for any $t \in \R$;
\item $\mbox{Span}(\cD)$ is dense in the Sobolev space $H^s(\R^3)$. This means that any function $W \in H^s(\R^3)$ can be approximated with an arbitrary accuracy $\epsilon > 0$ by a finite linear combination of functions of $\cD$, and therefore by a finite sum of functions of $\cD$ since $\cD$ is a cone: for any $\epsilon > 0$, there exists a finite integer $p \in \N^\ast$, and $p$ functions $\phi_1^{(p)}$, ... $\phi_p^{(p)}$ in $\cD$ such that 
$$
 \left\| W - \left( \sum_{j=1}^{p} \phi_j^{(p)}  \right) \right\|_{H^s} \le \epsilon.
$$
Greedy algorithms provide practical ways to construct such approximations;
\item $\cD$ is weakly closed in $H^s(\R^3)$. This technical assumption ensures the convergence of the greedy algorithm~\cite{temlyakov_rev}.
\end{enumerate}
Given a dictionary $\cD$, the greedy method then consists of
\begin{itemize}
\item initializing the algorithm with (for instance) $\widetilde W_0=0$;
\item constructing iteratively a sequence $\widetilde W_1, \widetilde W_2, \widetilde W_3, \cdots$ of more accurate approximations of the target Wannier function $W$ of the form
\begin{equation}\label{eq:defWk}
\widetilde W_p(\br) = \sum_{j=1}^p \phi_j^{(p)}(\br),
\end{equation}
where $\phi_j^{(p)}$ are functions of the dictionary $\cD$;
\item stopping the iterative process when $\|W-\widetilde W_p\|_{H^s} \le \epsilon$, where $\epsilon > 0$ is the desired accuracy (for the chosen $H^s$-norm).
\end{itemize}
We will use here the orthogonal greedy algorithm for constructing $\widetilde W_{p+1}$ from $\widetilde W_p$, which is defined as follows:

\medskip

\begin{algorithm}[Orthogonal greedy algorithm] $\,$
\label{alg:greedy}
\begin{description}
\item[Step 1:] Compute the residual at iteration $k$: 
$$
R_p(\br) =  W(\br) - \widetilde W_p(\br);
$$
\item[Step 2:]  find a local minimizer $\phi_{p+1}$ to the optimization problem
\begin{equation}\label{eq:greedy_outer}
\min_{\phi \in \cD} J_p(\phi), \quad \mbox{where} \quad J_p(\phi) := \|R_p-\phi\|_{H^s}^2;
\end{equation}
\item[Step 3:] solve the unconstrained quadratic optimization problem
\begin{equation} \label{eq:orth_greedy}
(c_j^{(p+1)})_{1 \le j \le p+1} \in  \mathop{\mbox{\rm argmin}} \left\{ \left\| W - \left( \sum_{j=1}^{p+1} c_j \phi_j^{(p)} + c_{p+1} \phi_{p+1} \right) \right\|_{H^s}^2, \; (c_j)_{1 \le j \le p+1} \in \R^{p+1} \right\};
\end{equation}
\item[Step 4:] set $\phi_j^{(p+1)} = c_j^{(p+1)} \phi_j^{(p)}$, $1 \le j \le p$, and $\phi_{p+1}^{(p+1)}=c_{p+1}^{(p+1)}  \phi_{p+1}$.
\end{description}
\end{algorithm}

Note that Step 3 is easy to perform since \eqref{eq:orth_greedy} is nothing but a least square problem in dimension $(p+1)$ ($p$ is of the order of $10$ to $10^3$ in practice). Step 2 will be described in detail in Sections~\ref{sec:alg_greedy} and~\ref{sec:alg_det}. The next section is concerned with the choice of the dictionary $\cD$.

%%%%%%%%%%%%%%%%%%%%%%%%
\subsection{Symmetry-adapted Wannier functions and Gaussian-type orbitals}
\label{sec:SAWF}

\medskip

We assume that we are dealing with a periodic material with space group $G = \cR \rtimes G_{\rm p}$,  where $\cR$ is a Bravais lattice embedded in $\R^3$, and $G_{\rm p}$ a finite point group (a finite subgroup of the orthogonal group $O(3)$). The Bravais lattice $\cR$ is two-dimensional for 2D materials such as graphene or hBN, and three-dimensional for usual 3D crystals. We also assume that we are given a symmetry-adapted Wannier function $W$ centered at a high-symmetry point $\bq \in \R^3$ of the crystalline lattice, and corresponding to a one-dimensional representation of the subgroup
$$
G_\bq^0:=\left\{ \Theta \in G_{\rm p} \; | \; \Theta \bq \in \bq + \cR \right\}
$$
of $G_{\rm p}$. For completeness, we include the basics of the theory of symmetry-adapted Wannier functions in the Appendix. Note that our method can straightforwardly be extended to the case of two-dimensional irreducible representations of $G_\bq^0$. We now translate the origin of the Cartesian frame to point $\bq$. Setting $G^0:=G_\bq^0$ to simplify the notation, the function $W$ satisfies in this new frame the invariance property
\begin{equation}
\label{eq:sym_adapt}
\forall \Theta \in G^0, \quad (\Theta W)(\br) = W(\Theta^{-1}\br) = \chi(\Theta) W(\br),
\end{equation}
where $\chi$ is the character of this one-dimensional representation.

\medskip
Our goal is to approximate the Wannier function $W$ by a finite sum of GTOs. In order to reduce the number of GTOs necessary to obtain the desired accuracy, while enforcing the symmetries of the approximate Wannier functions $\widetilde W_p$, we use a dictionary consisting of symmetry-adapted Gaussian-type orbitals (SAGTOs) of the form
\begin{equation}\label{eq:SAGTO}
\phi_{\balpha,\sigma,\Lambda}^{\rm SA}(\br) =\frac 1{|G^0|} \sum_{\Theta \in G^0} \chi(\Theta) \, (\Theta \varphi_{\balpha,\sigma,\Lambda})(\br) =  \frac 1{|G^0|} \sum_{\Theta \in G^0} \chi(\Theta) \, \varphi_{\balpha,\sigma,\Lambda}(\Theta^{-1}\br),
\end{equation}
where $|G^0|$ is the order of the group $G^0$, and
$$
\varphi_{\balpha,\sigma,\Lambda}(\br) = \left( \sum \limits_{(n_x,n_y,n_z) \in \cI} \lambda_{n_x,n_y,n_z} (r_x-\alpha_x)^{n_x} (r_y-\alpha_y)^{n_y} (r_z-\alpha_z)^{n_z} \right)  \exp\left( - \frac{1}{2 \sigma^2}  |\br- \balpha|^2 \right)
$$
is a Gaussian-polynomial function centered at $\balpha \in \R^3$ with standard deviation $\sigma > 0$. The set $\cI$ is a carefully chosen subset of 
$\left\{ (n_x,n_y,n_z) \in \N^3 \; | \; n_x+n_y+n_z \le L \right\}$ (total degree lower than or equal to $L$) determined by the symmetries of the SAWF. Note that for 2D materials on the $xy$ plane, it is more appropriate to chose $\cI \subset \left\{ (n_x,n_y,n_z) \in \N^3 \; | \; n_x+n_y \le L_\parallel, \; n_z \le L_\perp \right\}$. Any function $\phi_{\balpha,\sigma,\Lambda}^{\rm SA}$ of the dictionary thus satisfies the same symmetry property
$$
\forall \Theta \in G^0, \quad (\Theta \phi_{\balpha,\sigma,\Lambda}^{\rm SA})(\br) = \phi_{\balpha,\sigma,\Lambda}^{\rm SA}(\Theta^{-1}\br) = \chi(\Theta) \phi_{\balpha,\sigma,\Lambda}^{\rm SA}(\br)
$$
as the Wannier function $W$ to be approximated. 

%%%%%%%%%%%%%%%%%%%%%%%%
\subsection{A greedy algorithm for compressing SAWF into SAGTO}
\label{sec:alg_greedy}
It can be shown that the set
\begin{equation}\label{eq:dict}
\cD^{\rm SA} := \left\{ \phi_{\balpha,\sigma,\Lambda}^{\rm SA}, \; \balpha \in \R^3, \; \sigma \in [\sigma_{\rm min},\sigma_{\rm max}], \; \Lambda \in \R^{\cI_\balpha} \right\},
\end{equation}
where $0 < \sigma_{\rm min} < \sigma_{\rm max} < \infty$ are given parameters (chosen by the user), and $\cI_{\balpha}$ is a carefully chosen nonempty subset of $\N^3$ depending on the center $\balpha$ of the SAGTO,
%\begin{align*}
%\cI& \subset \left\{ (n_x,n_y,n_z) \in \N^3 \; | \; n_x+n_y+n_z \le L \right\} \; \mbox{ (3D material)}, \\ 
%\cI& \subset \left\{ (n_x,n_y,n_z) \in \N^3 \; | \; n_x+n_y \le L_\parallel, \; n_z \le L_\perp \right\}  \; \mbox{ (2D material)},
%\end{align*}
is a dictionary for the closed subspace 
$$
H^{s,{\rm SA}}(\R^3):= \left\{ f \in H^s(\R^3) \; | \; \forall \Theta \in G^0, \; (\Theta f)(\br) = f(\Theta^{-1}\br) = \chi(\Theta) f(\br) \right\}
$$
of $H^s(\R^3)$ for any $s \in \R_+$. 
For example, in the case of Graphene and hBN (see Section~\ref{sec:num}), we use the same set for each $\balpha \in \R^3$:
$$\cI_\balpha = \{(0,0,1), (0,0,3), (0,0,5)\}, \quad  \forall \balpha \in \R^3. $$ 
More refine strategies will be considered in future works. 
 
\medskip

The main practical difficulty in Algorithm~\ref{alg:greedy} is the computation of a local minimum to Problem~\eqref{eq:greedy_outer}. This problem can be formulated as 
\begin{equation}\label{eq:greedy_pb}
\min_{\balpha \in \R^3, \; \sigma \in [\sigma_{\rm min},\sigma_{\rm max}], \; \Lambda \in \R^\cI } \cJ_p(\balpha,\sigma,\Lambda), \quad \mbox{where} \quad \cJ_p(\balpha,\sigma,\Lambda) := \|R_p-\phi_{\balpha,\sigma,\Lambda}\|_{H^s}^2.
\end{equation}
The above minimization problem can in turn be written as:
\begin{equation}\label{eq:greedy_pb2}
\min_{\balpha \in \R^3, \; \sigma \in [\sigma_{\rm min},\sigma_{\rm max}]}   \widetilde\cJ_p(\balpha,\sigma),
\end{equation}
where 
\begin{equation}\label{eq:greedy_pb3}
\widetilde\cJ_p(\balpha,\sigma)=\min_{\Lambda \in \R^\cI } \cJ_p(\balpha,\sigma,\Lambda).
\end{equation}
Since the map $\Lambda \mapsto \cJ_p(\balpha,\sigma,\Lambda)$ is quadratic in $\Lambda$, problem~\eqref{eq:greedy_pb3} can be solved explicitly at a very low computational cost, and the gradient of  $\widetilde\cJ_p(\balpha,\sigma)$ with respect to both $\balpha$ and $\sigma$ can be easily computed from the solution of problem~\eqref{eq:greedy_pb3} by the chain rule. We can then use a constrained optimization solver to find a local minimizer to the four-dimensional optimization problem~\eqref{eq:greedy_pb2}.

%%%%%%%%%%%%%%%%%%%%%%%%
\subsection{Algorithmic details}
\label{sec:alg_det}

\subsubsection{Construction of MLWFs}
\label{sec:numerical_MLWF}

The Bloch energy bands and wave-functions of the periodic Kohn-Sham Hamiltonian are obtained using VASP with pseudo-potentials of the Projector Augmented Wave (PAW) type~\cite{blochl1994projector}, the PBE exchange-correlation functional~\cite{PBE96}, a plane-wave energy cutoff $E_{\rm c}$ and a grid $\cQ$ of the Brillouin zone $\Gamma^*$. For 2D materials, the height $\eta$ of the supercell is chosen sufficiently large to eliminate the spurious interactions between the material and its periodic images. The Bloch eigenfunctions belonging to the energy bands of interest are combined into a basis of MLWFs using the Marzari-Vanderbilt algorithm~\cite{MV97} as implemented in the Wannier90 computer program \cite{W90_code}. 
The final output is a set of Wannier functions which are known to be localized at a certain point and exponentially decaying for materials with suitable topological properties such as the ones considered in Section~\ref{sec:num} (see \cite{panati2013bloch}).
Using a sufficiently large rectangular box,
\[
    \Omega := [x_{\rm min}, x_{\rm max}] \times [y_{\rm min}, y_{\rm max}] \times [z_{\rm min}, z_{\rm max}] \subset \R^3,
\] 
we can neglect the exponentially vanishing values of the Wannier function under consideration outside the box. 
The numerical values of the Wannier function $W$ are given on a Cartesian grid $\cM$ spanning the box and containing $M=M_x M_y M_z$ points. The Wannier functions obtained in this manner are in general not perfectly symmetry-adapted, as the Marzari-Vanderbilt algorithm does not take symmetries into account. However, in practice, the MLWFs we generated are close enough to SAWFs so that it was possible to identify a high-symmetry center and an associated point group. To test our compression method, we symmetrize the MLWFs according to the identified point group before applying the greedy procedure.

%Thus, for each function $W$ centered at a high-symmetry point $\bq \in \R^3$, we identify the associated point group $G_\bq$, based on observed properties and on physical considerations.
%, and apply equation~\eqref{eq:sym_adapt} to enforce exactly the symmetries, yielding a SAWF. 

\subsubsection{Optimization Procedure in the Discrete Setting}
\label{sec:disc_greed}

We present next the discrete formulation of problem~\eqref{eq:greedy_pb3}. The discrete data representing the Wannier function $W$ centered at $\bq \in \R^3$ are composed of: i) the symmetry group $G^0$ and ii) the point values $(W(\br) ) _{\br \in \cM}$ at each point of the cartesian grid $\cM$. Because we seek to minimize in particular the $H^1$-norm of the residual, we introduce an auxiliary Fourier representation of the data. Indeed, computing gradients is a fast (diagonal) operation in momentum space. The Fast Fourier Transform algorithm (FFT) can be used to efficiently transform data from position to momentum space.
In particular, we obtain the unnormalized discrete representation of the Fourier transform $\widehat{u}$ of any function $u$ as point values $(\widehat{u}(\bk))_{\bk \in \cK}$ on a secondary Cartesian momentum-space grid that we denote by $\cK$, containing the same number of points as the real-space grid, i.e $|\cK| = |\cM| = M$. Let us recall that the FFT algorithm requires $M_x$, $M_y$ and $M_z$ to be even numbers so that the momentum grid $\cK$ is centered at zero. The $H^s$--norm~\eqref{eq:defHs} of $u$ then has a discrete approximation given by
\begin{equation}
\Vert u \Vert_{H^s}^2 \approx  \frac{\vert \Omega \vert}{M^2} \sum_{\bk \in \cK}   \left (1 + \vert \bk \vert^2 \right )^s \left \vert \widehat{u}(\bk)\right \vert^2 . 
\end{equation}
At every greedy iteration $p \ge 0$, the exact cost functional $\cJ_p$ is approximated in the discrete setting by the functional $\cJ^\cM_{p}$ defined as:
\begin{equation}
\label{eq:def_J_p_discrete}
\cJ^\cM_{p}(\balpha, \sigma, \Lambda) := \frac{\vert\Omega\vert}{M^2}   \sum_{\bk \in \cK}   \left (1 + \vert \bk \vert^2 \right )^s \left \vert \widehat{R}_p(\bk) -  \widehat{\phi}_{\balpha, \sigma, \Lambda}(\bk) \right\vert^2 ,
\end{equation}
where we recall that the residual $R_p$ is computed from the approximation $\widetilde{W_p}$ at step $p$ of the target Wannier function $W$,
\[
    R_p(\br) = W(\br) - \widetilde{W_p}(\br).
\]
Note that while the Fourier transform of the SAGTO function $\phi_{\balpha, \sigma, \Lambda}$ which appears in this expression can be analytically computed, it is faster and more consistent to evaluate directly the Fourier transform of the residual numerically using the FFT algorithm.

\medskip 

For the implementation of the minimization problem~\eqref{eq:greedy_pb2} with the discrete error functional~\eqref{eq:def_J_p_discrete}, we use a constrained optimization solver to find a local minimizer to the non-convex minimization problem
\begin{equation}\label{eq:discrete_greedy_pb2}
\min_{\balpha \in \Omega, \; \sigma \in [\sigma_{\rm min},\sigma_{\rm max}]}   \widetilde\cJ^\cM_p(\balpha,\sigma),
\end{equation}
the minimization over the coefficients $\Lambda$ of the SAGTO being performed explicitly for fixed $\balpha,\ \sigma$ by solving the least-square problem
\begin{equation}\label{eq:discrete_greedy_pb3}
\widetilde\cJ^\cM_p(\balpha,\sigma)=\min_{\Lambda \in \R^\cI } \cJ^\cM_p(\balpha,\sigma,\Lambda).
\end{equation}
We tested both the \emph{Sequential Quadratic Programming} (SQP) and the \emph{Interior-Point} (IP) specializations of the \emph{fmincon} optimization routine implemented in the Matlab Optimization Toolbox \cite{MATLAB}. 
To accelerate the computation, the gradient (but not the Hessian matrix) is also provided to the optimizer routine. 
Note that it is straightforward to compute explicitly the gradient by the chain rule in the case of the discrete error functional in~\eqref{eq:discrete_greedy_pb2} from the solution of the inner problem in~\eqref{eq:discrete_greedy_pb3} (its expression is quite cumbersome). The iterative procedure is stopped when one of the following two convergence criteria is met: (i) the norm of the gradient is smaller than $\delta = 10^{-10}$; (ii) the relative step size between two successive iterations is smaller than $\tau_{\rm min}= 10^{-12}$. 
In practice, our numerical tests show that both optimization routines (SQP or IP) provide similar results, with the IP method being slightly faster.

As usual with non-convex optimization problems, it is very important to provide a suitable initial guess for the parameters, in the present case the center of the Gaussian $\balpha^0 \in \Omega$ and its variance $\sigma_{\rm min} \leq \sigma^0 \leq \sigma_{\rm max}$. We propose here the following initialization procedure.
First, the initial center position $\balpha^0$ is chosen as a maximizer of the absolute value of the residual $R_p$: 
\begin{equation}
\label{eq:guess_alpha}
\balpha^0 \in  \mathop{\rm argmax} \limits_{\br \in \Omega} |R_p(\br) |.
\end{equation} Next, two different heuristic guesses are proposed to determine a suitable initial value $\sigma^0$, assuming that the function $\vert R_p \vert$ resembles locally a Gaussian function centered at $\balpha^0$,
\begin{equation}\label{eq:local_behavior}
    |R_p(\br)|  \approx |R_p(\balpha^0)| \exp\left(-\frac{\vert \br - \balpha^0\vert^2}{2\sigma^2}  \right).
\end{equation}
A first guess for $\sigma^0$ is obtained by a local data fit,
\[
    \sigma^0_1 = \mathop{\rm argmin} \limits_{\sigma > 0} \sum_{\br \in \cM \cap B(\balpha^0)} \left ( \frac{1}{2\sigma^2} \left \vert \br - \balpha^0 \right \vert^2\ +\ \log \left \vert \frac{ R_p(\br) }{R_p(\balpha^0)} \right \vert \right )^2,
\] 
where $B(\balpha^0)$ is a cubic box centered at $\balpha^0$ of side length $2r_{\rm cutoff}$, with $r_{\rm cutoff}$ a user-defined parameter.
This is in fact a linear least-squares fit, yielding the explicit formula:
\begin{equation}
\label{eq:guess_sigma_loc}
\sigma_1^0 = \left(\frac{ \dps \sum_{\br \in \cM \cap B(\balpha^0)} \left \vert \br - \balpha^0 \right \vert^4 }{-2 \dps\sum_{\br \in \cM \cap B(\balpha^0)} \left \vert \br - \balpha^0 \right \vert^2 \log \left \vert \frac{ R_p(\br) }{R_p(\balpha^0)} \right \vert}  \right)^{1/2}.
\end{equation} A second guess is provided by a property linking the variance of the standard normalized Gaussian $g(\br) = (2\pi\sigma^2)^{-1/2} \exp\left( - \frac{1}{2 \sigma^2}  \vert \br \vert^2 \right)$ to its \emph{full width at half maximum}, denoted $\omega_{\rm h}$:
\[
    \frac{\omega_{\rm h}[g]}{\sigma} = 2 \sqrt{2 \log 2}.
\]
The full width at half maximum is not well defined for arbitrary (non-radial) functions. We choose here to sample the full-width at half maximum along one-dimensional slices in all three directions $x,y,z$ around $\balpha^0$ and retain the smallest value. For an arbitrary function $u$ assumed to have its maximum magnitude at the origin, we let: 
\[
\omega_h[u] :=  \min_{d \in \{ x,y,z \}}\inf \left \{ \left \vert \gamma_+ - \gamma_- \right \vert : \qquad \gamma_- < 0 < \gamma_+ \text{ and } \left \vert \frac{ u \left (\gamma_{\pm} \mathbf{e}_d\right )}{u \left (\boldsymbol{0} \right )}\right  \vert \leq \frac 12 \right \},
\]
where $\mathbf{e}_d$ is the standard unit vector in the direction $d \in \{x,y,z\}$. This leads to a second initial guess for the variance:
\begin{equation}
\label{eq:guess_sigma_half}
\sigma_{2}^0 = \frac{ \omega_{\rm h}\left [ R_p(\cdot - \balpha^0)  \right ]}{ 2 \sqrt{2 \log 2}}.
\end{equation} In practice, we project the values $\sigma^0_{1}$ given by~\eqref{eq:guess_sigma_loc} and $\sigma^0_{2}$ given by~\eqref{eq:guess_sigma_half} on the interval $[\sigma_{\rm min}, \sigma_{\rm max}]$ and choose 
\begin{equation}
\sigma^0 = \mathop{\rm argmin} \limits_{i=1,2} \cJ_p(\balpha^0, \sigma^0_i, \Lambda^0).
\end{equation} Again, we do not claim that this procedure is optimal; it however gives satisfactory results for all the test cases we ran.

%%%%%%%%%%%%%%%%%%%%%%%%
\section{Numerical results}
\label{sec:num}
Our greedy algorithm allows us to compress a SAWF defined on a cartesian grid with $M=M_xM_yM_z$ points into a sum of SAGTOs parameterized by $p(4+|\cI|)$ real numbers, where $p$ is the number of SAGTOs in the expansion
$$
\widetilde W_p^{\rm SA}(\br) = \sum_{j=1}^p \phi^{\rm SA}_{\balpha_j,\sigma_j,\Lambda_j}(\br),
$$
and where each $\phi^{\rm SA}_{\balpha_j,\sigma_j,\Lambda_j}$ is characterized by $(4+|\cI|)$ real parameters.
The compression gains for the four numerical examples detailed below, namely three 2D materials (single-layer graphene, hBN, and FeSe), and one bulk crystal (diamond-phase silicon), are collected in Table~\ref{tab:original_compressed}.  The numerical parameters used in the construction of the original Wannier functions (as described in Section~\ref{sec:numerical_MLWF}) are given in Table~\ref{tab:data}. 
\begin{table}[!htb]
\centering
\begin{tabular}{|c|c|c|c|c|c|c|} 
\hline
 Material 									& $M$  											    & $|\cI|$ 	 						     &$\epsilon$						&$p$      		&$p(4+|\cI|)$	 	& Compression ratio \\ \hline \hline
\multirow{3}{*}{Graphene} 	& \multirow{3}{*}{$3237696$}		&\multirow{3}{*}{$3$}	     & $0.1$  							&$115$  		& $805$	  		&$4022$ \\  
												&															&				 						     &	 $0.02$ 			 				& $1036$ 	& $7252$	  	&$446$ \\ \hline \hline 
\multirow{3}{*}{hBN} 			& \multirow{3}{*}{$4021248$}		&\multirow{3}{*}{$3$}	     & $ 0.1$  							&$137$  		&  $959$		&$4193$ \\  
												&															&				 						     &	 $0.03$ 			 				& $1500$  	 &  $10500$	&$383$ \\ \hline \hline 
\multirow{3}{*}{Si} 				& \multirow{3}{*}{$110592$}			&\multirow{3}{*}{$3$}	     & $0.1$  							& $424$  	&	 $2968$			&$38$ \\  
												&															&				 						     &	 $0.02$ 			 				& $1500$  	&  $10500$		&$10$ \\ \hline \hline 
%												&															&				 							 &	 $-$ 			 					& $-$  		&  $-$					&$-$ \\    \hline 
\multirow{3}{*}{FeSe} 			& \multirow{3}{*}{$4032000$}		&\multirow{3}{*}{$2$}	     & $0.1$  							&$133$  			&  $798$					&$5052$ \\  
												&															&				 						     &	 $0.02$ 			 				& $1610$  		&  $9660$					&$417$ \\ \hline \hline 
\end{tabular}
\caption{\label{tab:original_compressed} Compression gains obtained with our implementation of the orthogonal greedy minimizing the $H^1$-norm of the residual for Wannier functions of graphene, hBN, FeSe, and bulk silicon, for different tolerance levels~$\epsilon$.}
\end{table}
%%%%%%%%%%%%%%%%%%%%%%%%%%%%%%%%%%%%%%%%%%%%%%%%%%
%%%%%%%%%%%%%%%%%%%%%%%%%%%%%%%%%%%%%%%%%%%%%%%%%%
\begin{table}[!htb]
\centering
\begin{tabular}{|c|c|c|c|c|c|}
\hline 
Material 		& $E_{\rm c} [eV]$	& $\cQ$ 						& $\eta$ [\AA]  			  & $\cM$  \\ \hline
Graphene 	& $500$ 			& $25\times 25\times 1$		& $20$					  & $168 \times 132 \times 146$ \\
hBN		 	& $500$ 			& $25\times 25\times 1$		& $20$					  & $192 \times 154 \times 136$ \\
FeSe		 & $500$ 			& $19\times 19\times 1$		& $25$					  & $120 \times 120 \times 280$ \\
Si		 	& $300$ 			& $7\times 7\times 7$		& $-$					  & $48 \times 48 \times 48$ \\ \hline 
\end{tabular}
\caption{\label{tab:data} Numerical parameters used for the construction of the original Wannier functions using VASP and Wannier90.}
\end{table}

\subsection{Graphene and single-layer hBN}
\begin{figure}[h]
\centering
\begin{tabular}{cc}
\includegraphics[height=5truecm]{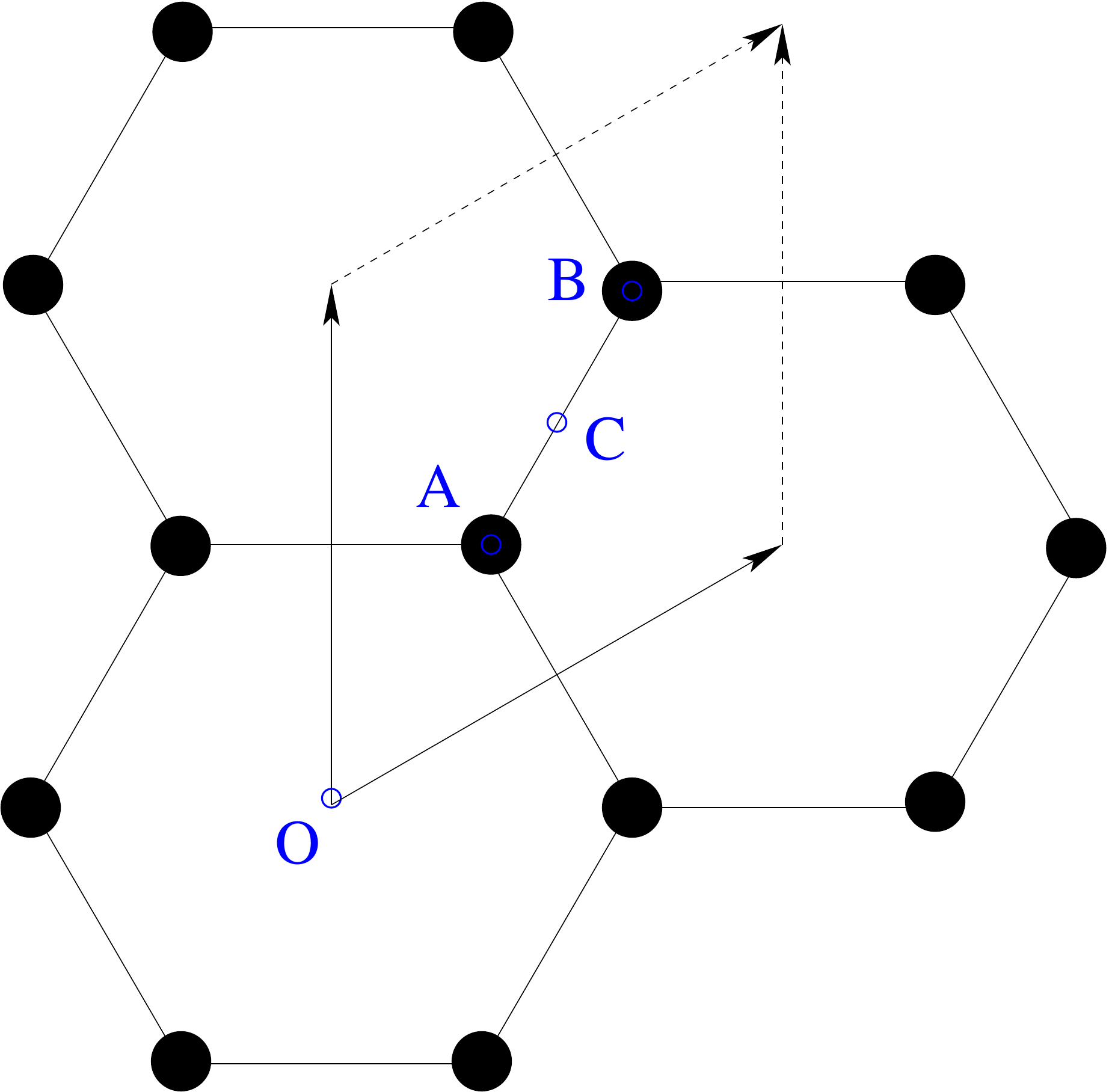} \qquad &  \qquad  \includegraphics[height=5truecm]{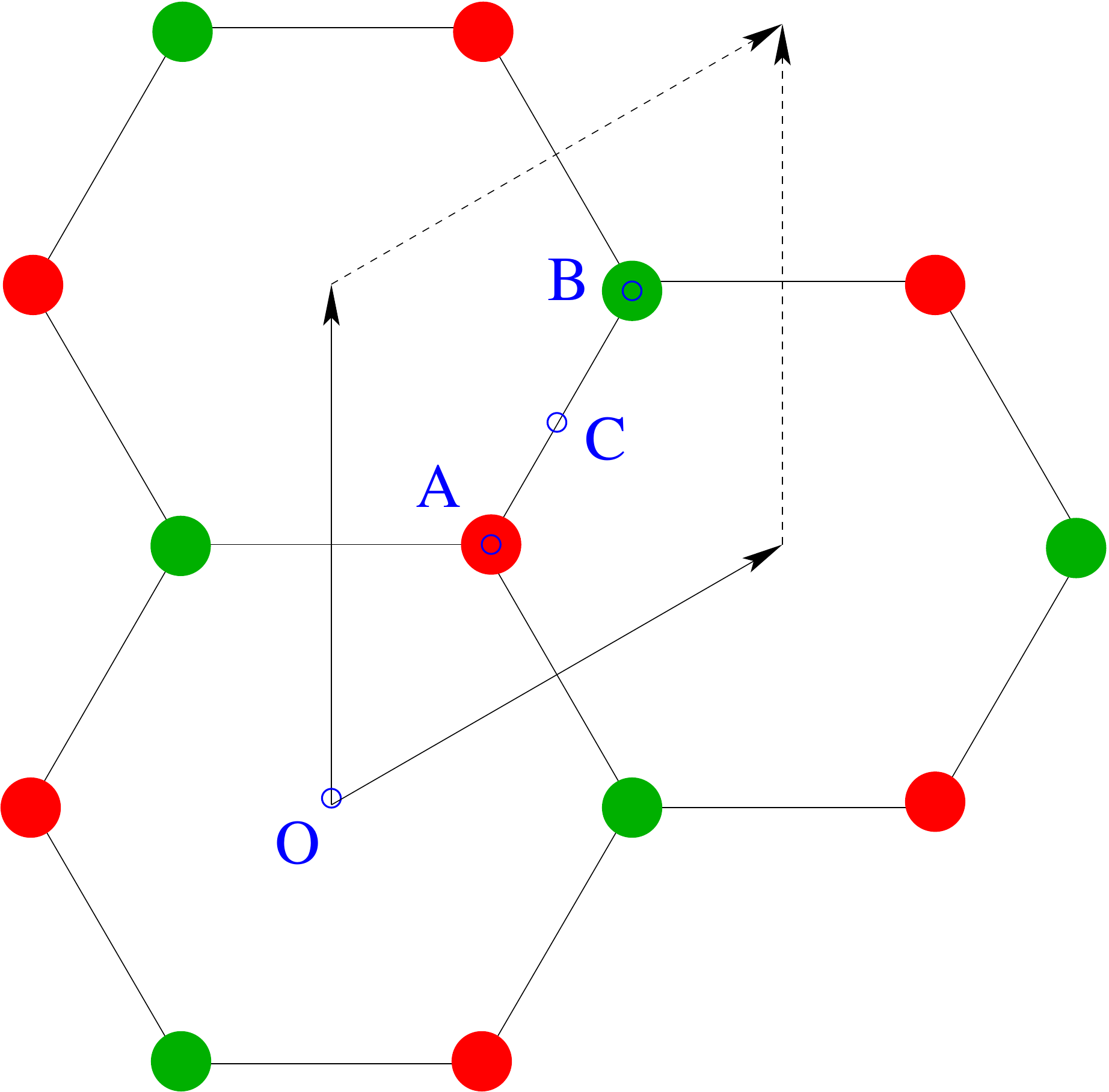} 
\end{tabular}
\caption{The honeycomb lattices of graphene (left) and hBN (right). The black dots represent carbon atoms, the red dots boron atoms, and the green dots nitrogen atoms. The blue dots $O$, $A$, $B$, and $C$ represent some high-symmetry points. \label{fig:graphene}}
\end{figure}

The space groups of graphene and single-layer hBN are respectively  
\begin{align*}
G &=\mbox{Dg80} := \cR \rtimes \underline{\rm D}_{6h}, \qquad \mbox{(space group of graphene)}, \\
G &=\mbox{P$\overline{6}$m2} := \cR \rtimes \underline{\rm D}_{3h}, \qquad \mbox{(space group of single-layer hBN)},
\end{align*}
where $\cR$ is the 2D Bravais lattice embedded in $\R^3$ defined as
\begin{equation}\label{eq:Bravais_graphene}
\cR = \Z a \left( \begin{array}{c} \sqrt 3/2 \\1/2 \\0 \end{array} \right) + \Z a \left( \begin{array}{c} 0 \\1 \\0 \end{array} \right), 
\end{equation}
where $a > 0$ is the lattice parameter (which takes different values for graphene and hBN). The group $\underline{\rm D}_{6h}$ is a group of order 24, and has 12 irreducible representations, while the group $\underline{\rm D}_{3h}$ is a group of order 12, and has 6 irreducible representations. The points $O$, $A$, $B$ and $C$ represented in Figure~\ref{fig:graphene} are high-symmetry points of graphene (left) and hBN (right); their symmetry groups are respectively
\begin{align*}
& G_O \equiv  \underline{D}_{6h}, \quad G_A \equiv \underline{D}_{3h}, \quad G_B \equiv \underline{D}_{3h}, \quad G_C \equiv  \underline{D}_{2h}, \quad \mbox{(graphene)}, \\
&G_O \equiv  \underline{D}_{3h}, \quad G_A \equiv \underline{D}_{3h}, \quad G_B \equiv \underline{D}_{3h}, \quad G_C \equiv  \underline{D}_{1h}, \quad \mbox{(single-layer hBN)}.
\end{align*}

Let $\sigma_h$ be the reflection operator with respect to the horizontal plane containing the graphene sheet. The two irreducible representations of the subgroup $\underline{C}_s=(E,\sigma_h)$ of $\underline{D}_{6h}$ and $\underline{D}_{3h}$ give rise to the decomposition of $L^2(\R^3)$ as
$$
L^2(\R^3) = L^2_+(\R^3) \oplus L^2_-(\R^3),
$$
where 
$$
L^2_+(\R^3) = \mbox{Ker}(\sigma_h-1), \quad L^2_-(\R^3) = \mbox{Ker}(\sigma_h+1).
$$
The bands associated with $L^2_+(\R^3)$ are the $\sigma$ bands, the ones associated with $L^2_-(\R^3)$ the $\pi$ bands. The bands of  interest for graphene and single-layer hBN are the valence and conduction bands closer to the Fermi level. For graphene, these are the $\pi$ bands originating from the 2p$_z$ orbitals of the carbon atoms.

The SAWF functions for graphene and single-layer hBN considered here are centered at point $A$ and are transformed according to the (one-dimensional) A$''_2$ representation of $\underline{D}_{3h}$, whose character is given in Table~\ref{tab:characters}.

\begin{figure}[!htb]
\centering
\subfloat{\includegraphics[width=7cm]{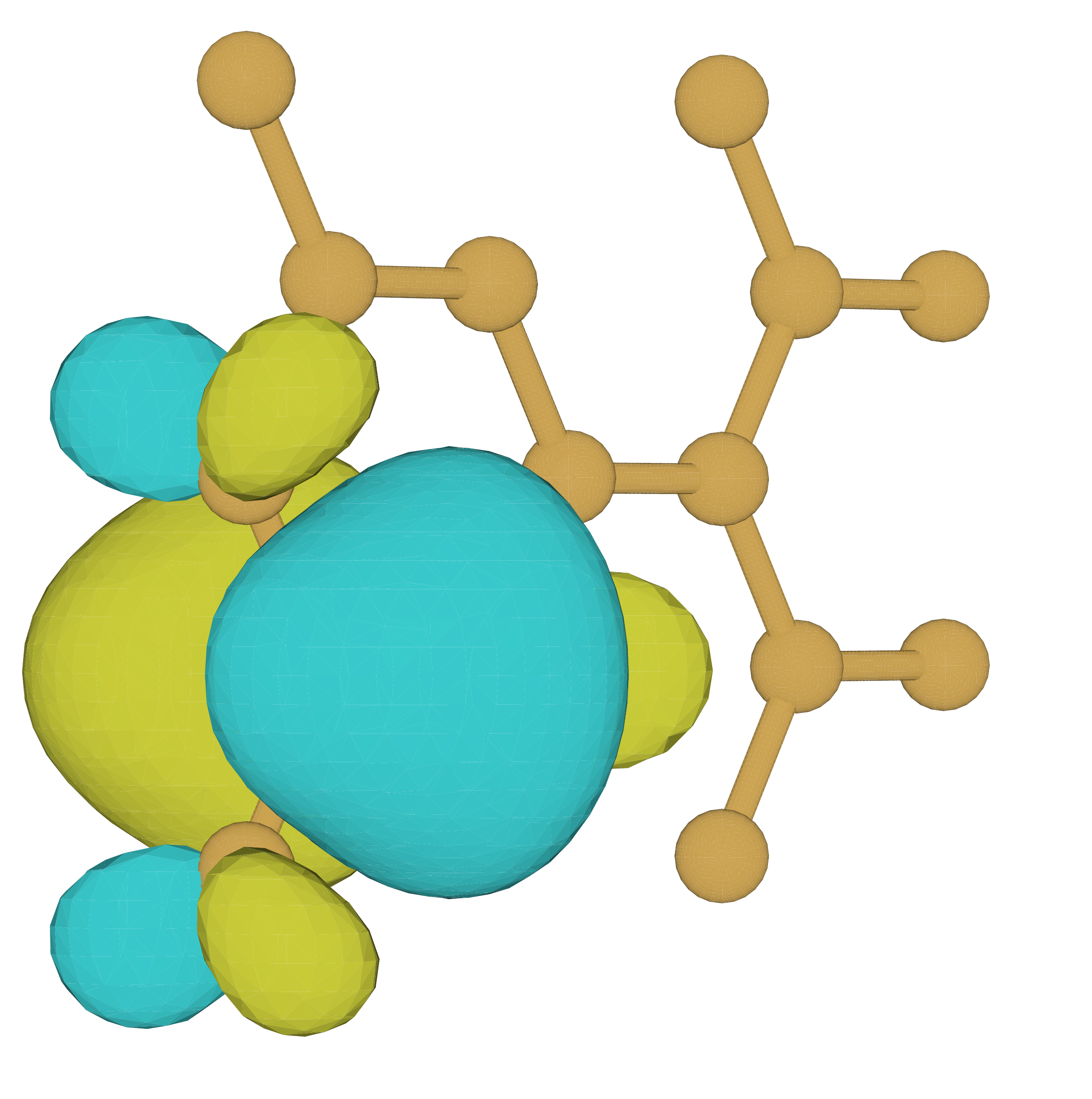}}
\subfloat{\includegraphics[width=7cm]{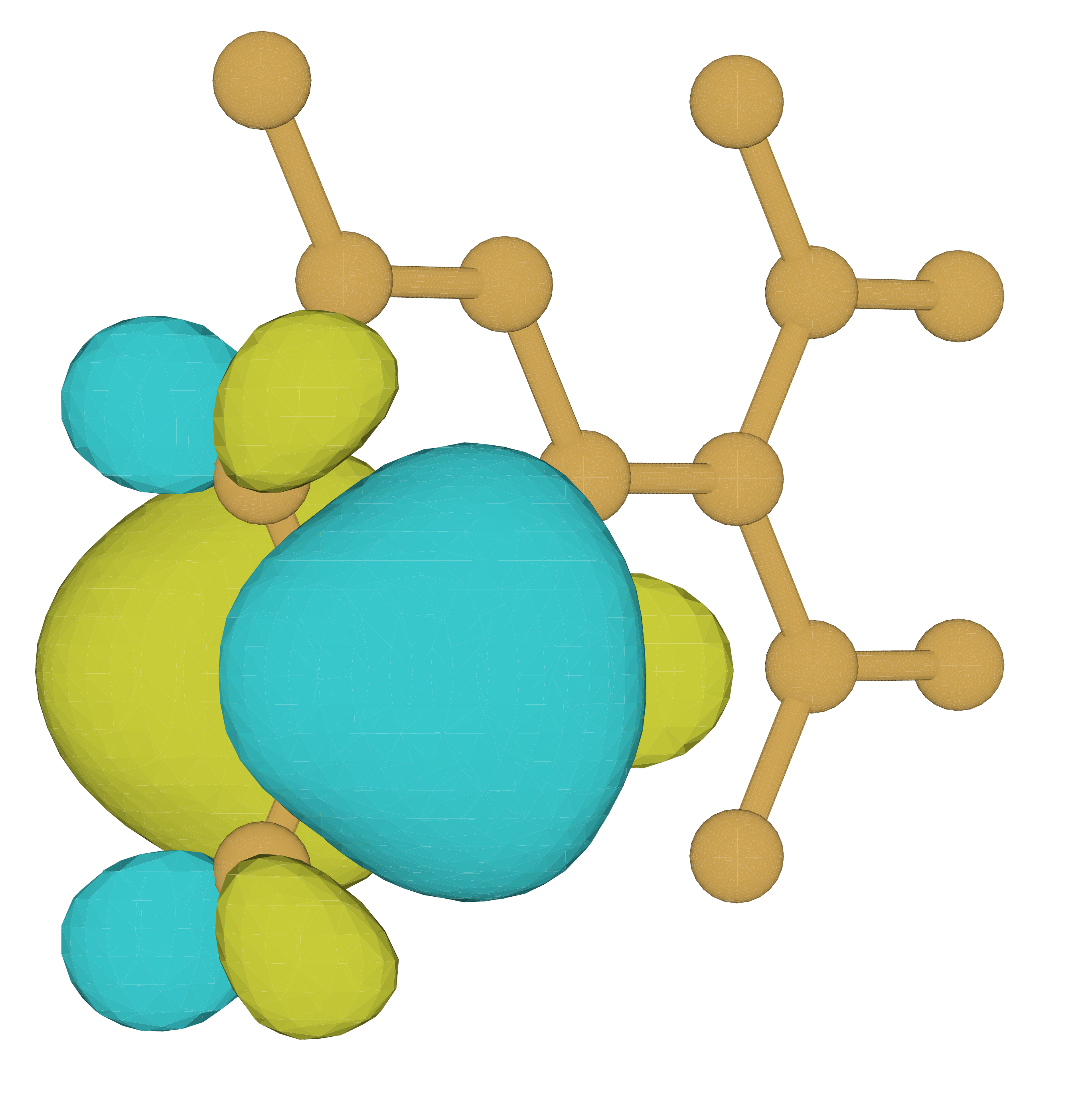}} 
\caption{\label{fig:Graphene} Wannier function of graphene generated with VASP and Wannier90 (left), and its compression into Gaussian orbitals (right). Positive and negative iso-surfaces corresponding to $15\%$ of the maximum value are plotted. \label{fig:WFgraphene}.}
\end{figure}

\begin{figure}[!htb]
\centering
\subfloat[]{\includegraphics[width=7cm]{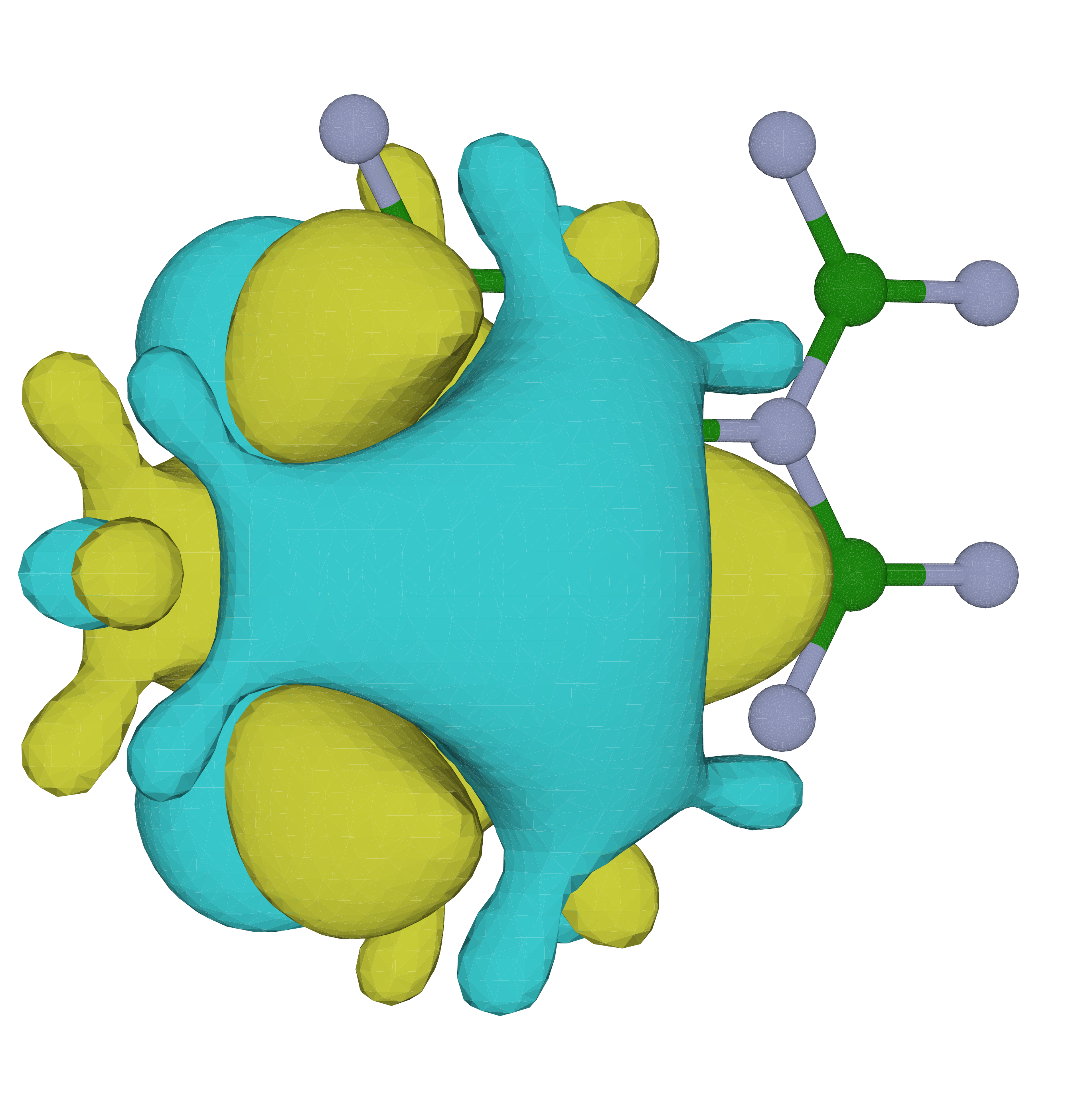}} 
\subfloat[]{\includegraphics[width=7cm]{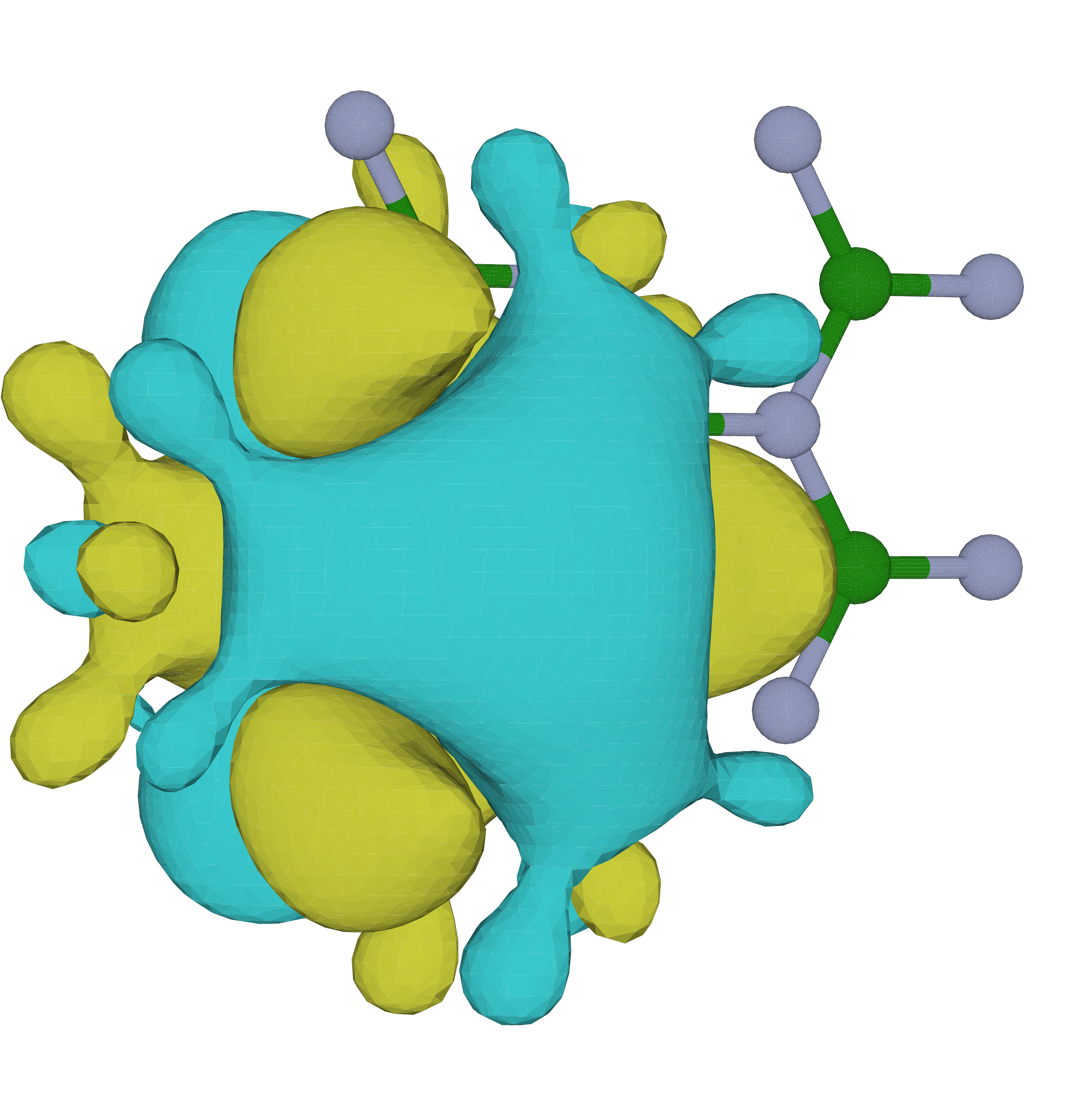}} 
\caption{\label{fig:hBN} Wannier function of single-layer hBN generated with VASP and Wannier90 (left), and its compression into Gaussian orbitals (right). Positive and negative iso-surfaces corresponding to $15\%$ of the maximum value are plotted. \label{fig:WFhBN}}
\end{figure}

\begin{figure}[!htb]
\centering
\subfloat{\includegraphics[width=0.5\linewidth]{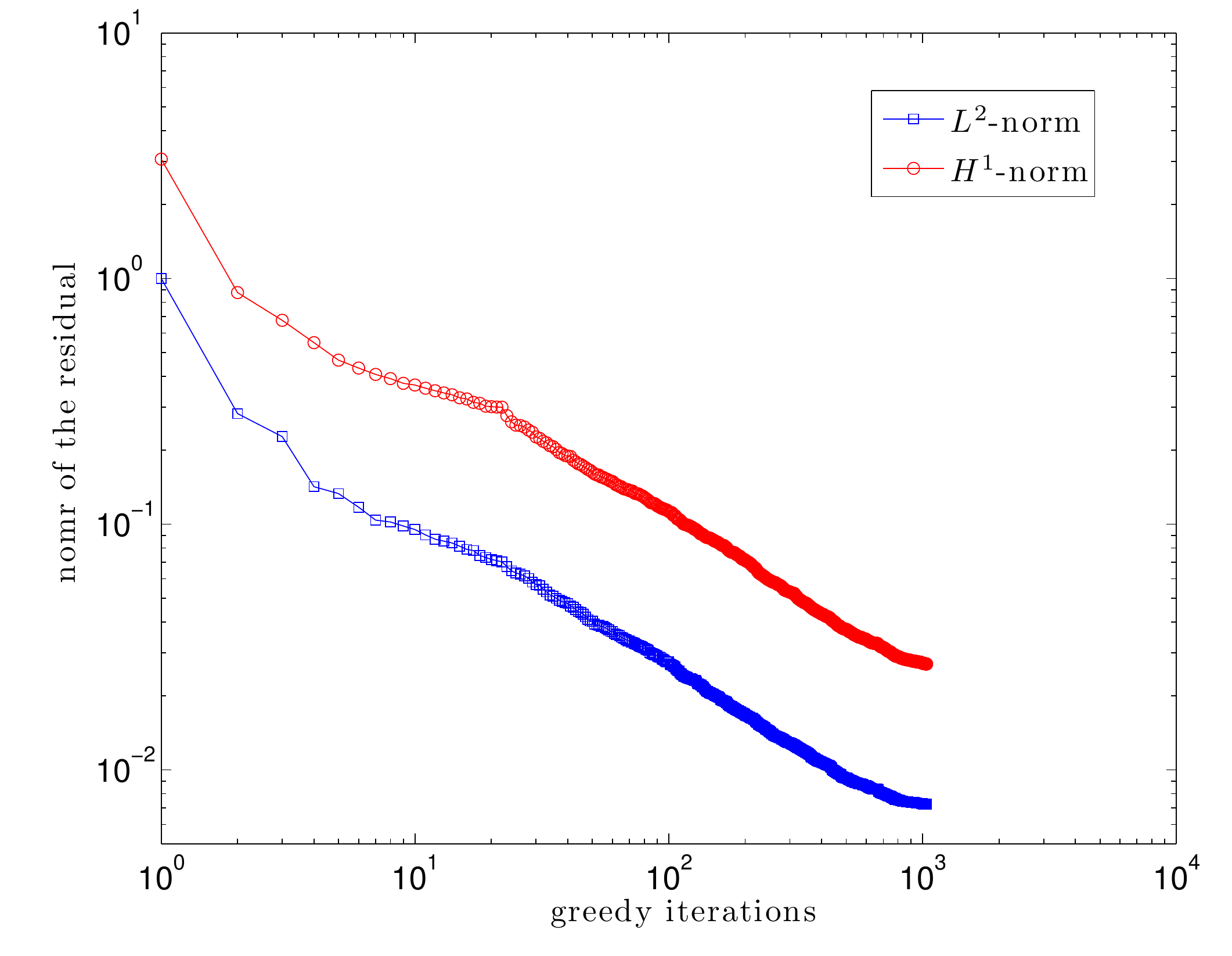}} 
\subfloat{\includegraphics[width=0.5\linewidth]{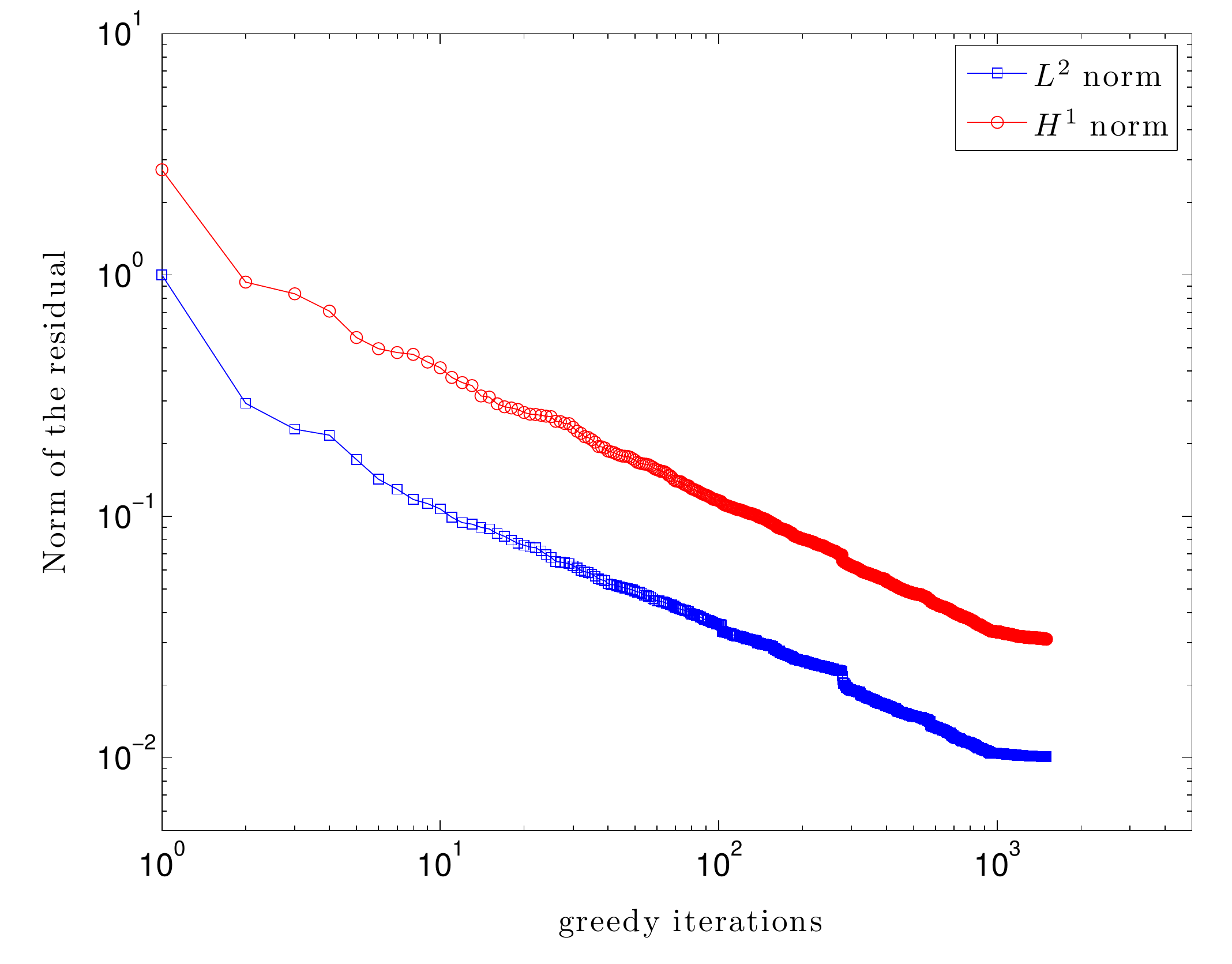}} 
\caption{\label{fig:Graphene_error} Decays of the $L^2$ and $H^1$-norms of the residual for our implementation of the orthogonal greedy algorithm minimizing the $H^1$-norm of the residual (left: graphene, right: hBN)\label{fig:residual_decay}}
\end{figure}

Graphical representations of the original Wannier functions generated by Wannier90 and of their compressions into Gaussian orbitals obtained with the VESTA visualization package~\cite{VESTA}, are displayed in Figures~\ref{fig:WFgraphene} (graphene) and \ref{fig:WFhBN} (hBN). The decays of the $L^2$ and $H^1$-norms of the residuals along the iterations of our implementation of the orthogonal greedy algorithm aiming at minimizing the $H^1$-norm of the residual, are plotted in Figure~\ref{fig:residual_decay}.

\begin{table}
\centering
\begin{tabular}{|c|c|c|c|c|c|c|c|c|c|}
\hline
$\underline{D}_{3h}$ & E	 & 2C$_3$ (z) & 3C$'_2$	& $\sigma_{\rm h}$ (xy) & 2S$_3$ & 3$\sigma_{\rm v}$ & linear  & quadratic  & cubic  \\	
& & & & & & & functions & functions & functions \\
\hline
A$''_2$ & +1 & +1 & -1 & -1 & -1 & +1 & $z$ & - & $z^3$, $z(x^2+y^2)$ \\
\hline
\end{tabular}
\caption{Character of the A$''_2$ representation of the group $\underline{D}_{3h}$ \label{tab:characters}}
\end{table}

%% Tim's suggestion 
%\begin{figure}[!htb]
%\centering
%\begin{tabular}{cc}
%\includegraphics[width=0.4\linewidth]{graphene_direct.pdf}  \quad & \quad
%\includegraphics[width=0.4\linewidth]{hBN_direct.pdf} \\
%\includegraphics[width=0.4\linewidth]{G_true_white.eps} \quad & \quad
%\includegraphics[width=0.4\linewidth]{hBN_true_white.eps} \\
%\includegraphics[width=0.4\linewidth]{G_compress_white.eps} \quad & \quad
%\includegraphics[width=0.4\linewidth]{hBN_compress_white.eps} \\
%\end{tabular}
%\caption{Tim's suggestion ! G and hBN }
%\end{figure}

\subsection{Single-layer SeFe}
\begin{figure}[!htb]
\centering
\subfloat[side view]{\includegraphics[width=4cm, height=5cm]{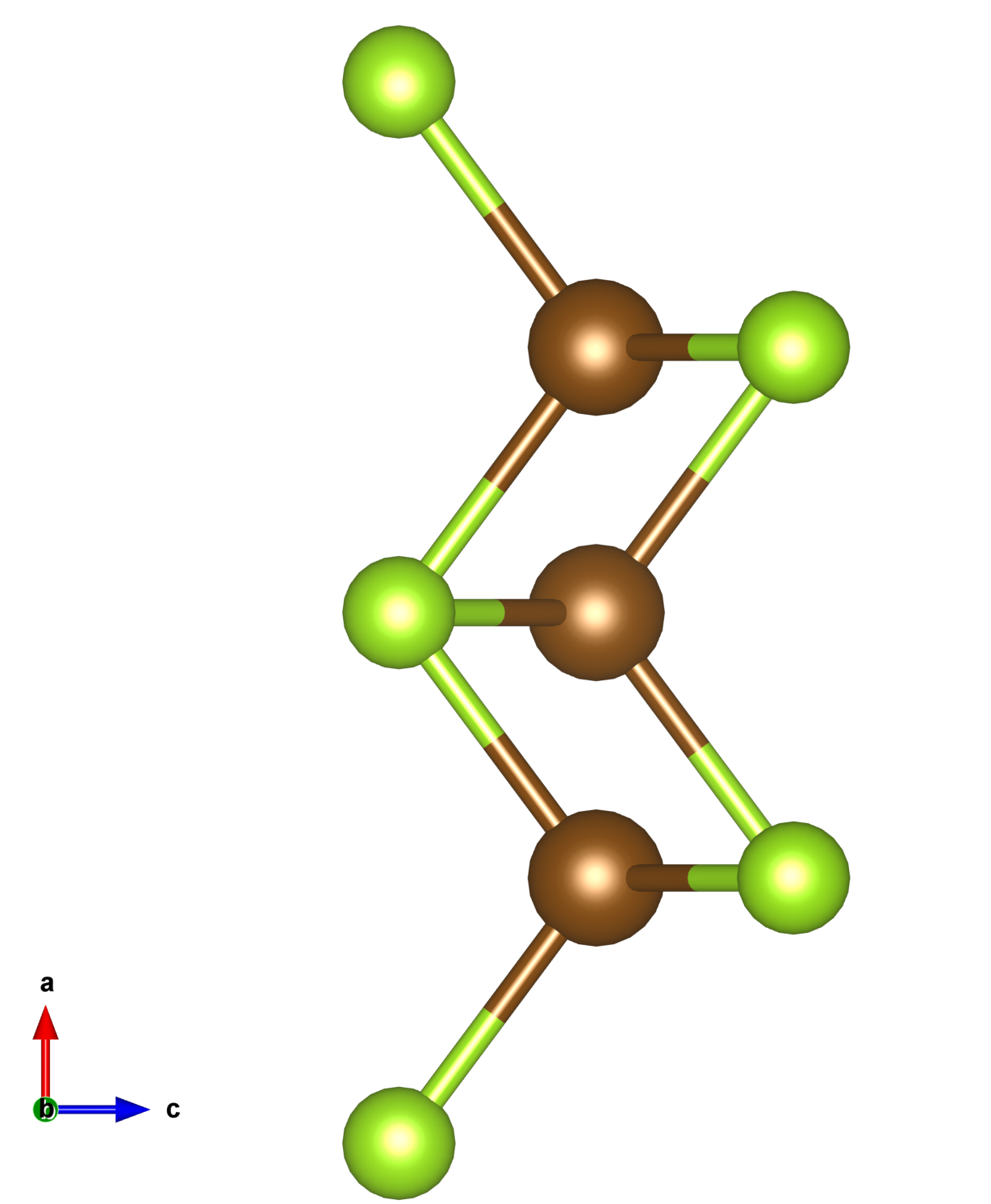}} \qquad 
\subfloat[top view]{\includegraphics[width=6cm, height=5cm]{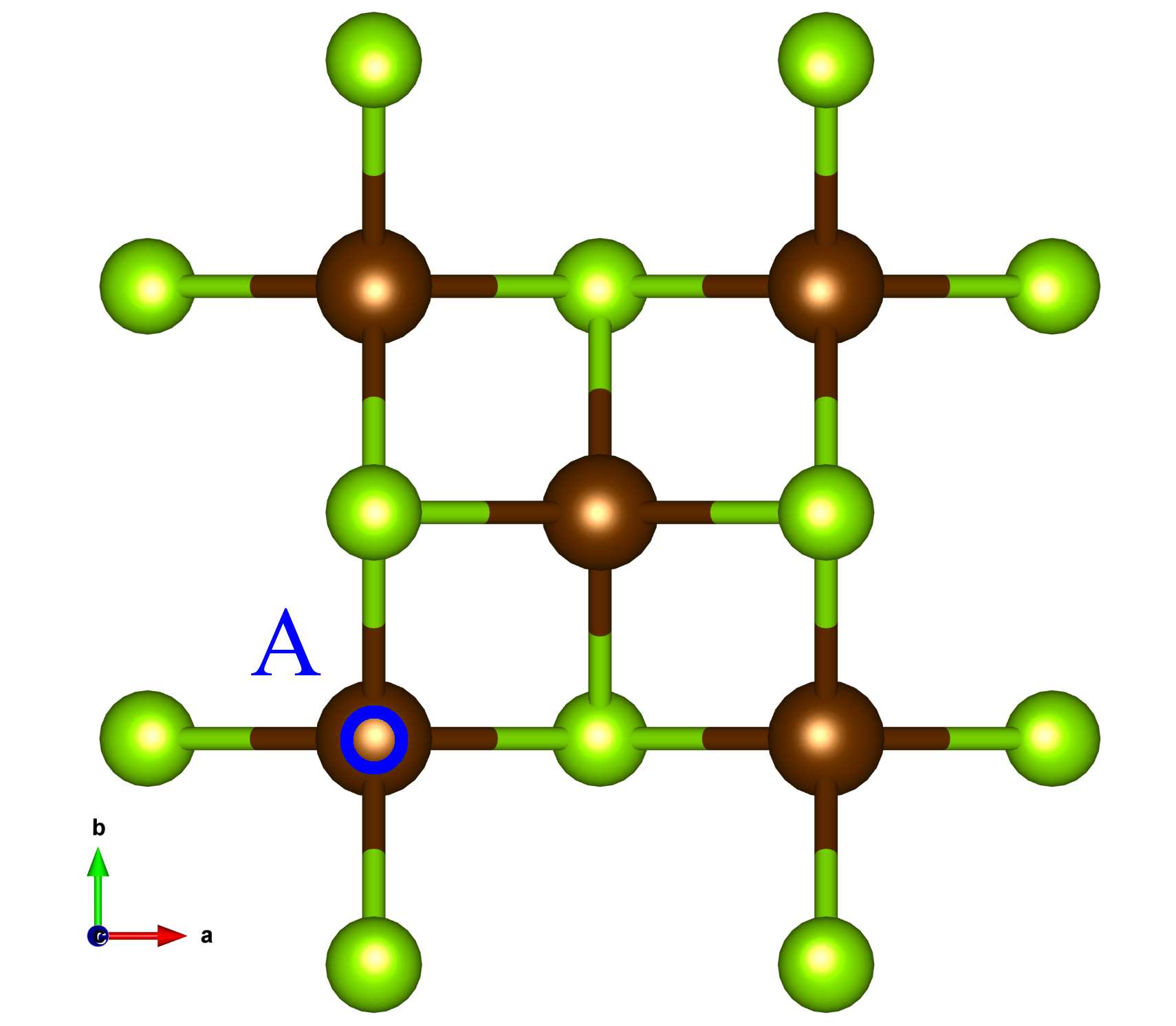}}
\caption{\label{fig:FeSe_sym} Crystalline structure of FeSe (2D layer with a finite thickness). The brown balls represent Fe atoms and the green balls represent Se atoms. The spotted point $A$ corresponds to the high-symmetry point at which the Wannier function is centered.} 
\end{figure}

The space group of single-layer FeSe is 
\begin{equation*}
G = \mbox{P4/nmm} := \cR \rtimes \underline{\rm D}_{4h},
\end{equation*}
where $\cR$ is the 2D square lattice of $\R^3$ defined as
\begin{equation}\label{eq:Bravais_graphene}
\cR = \Z a \left( \begin{array}{c} 1 \\0 \\0 \end{array} \right) + \Z a \left( \begin{array}{c} 0 \\1 \\0 \end{array} \right), 
\end{equation}
where $a > 0$ is the lattice parameter.  The group $\underline{D}_{4h}$ is of order 16 and has 10 irreducible representations. The symmetry group of the high-symmetry point $A$ represented in Figure~\ref{fig:FeSe_sym} is $G_A=\underline{C}_{2v}$.

\begin{figure}[!htb]
\centering
\subfloat[]{\includegraphics[width=7cm]{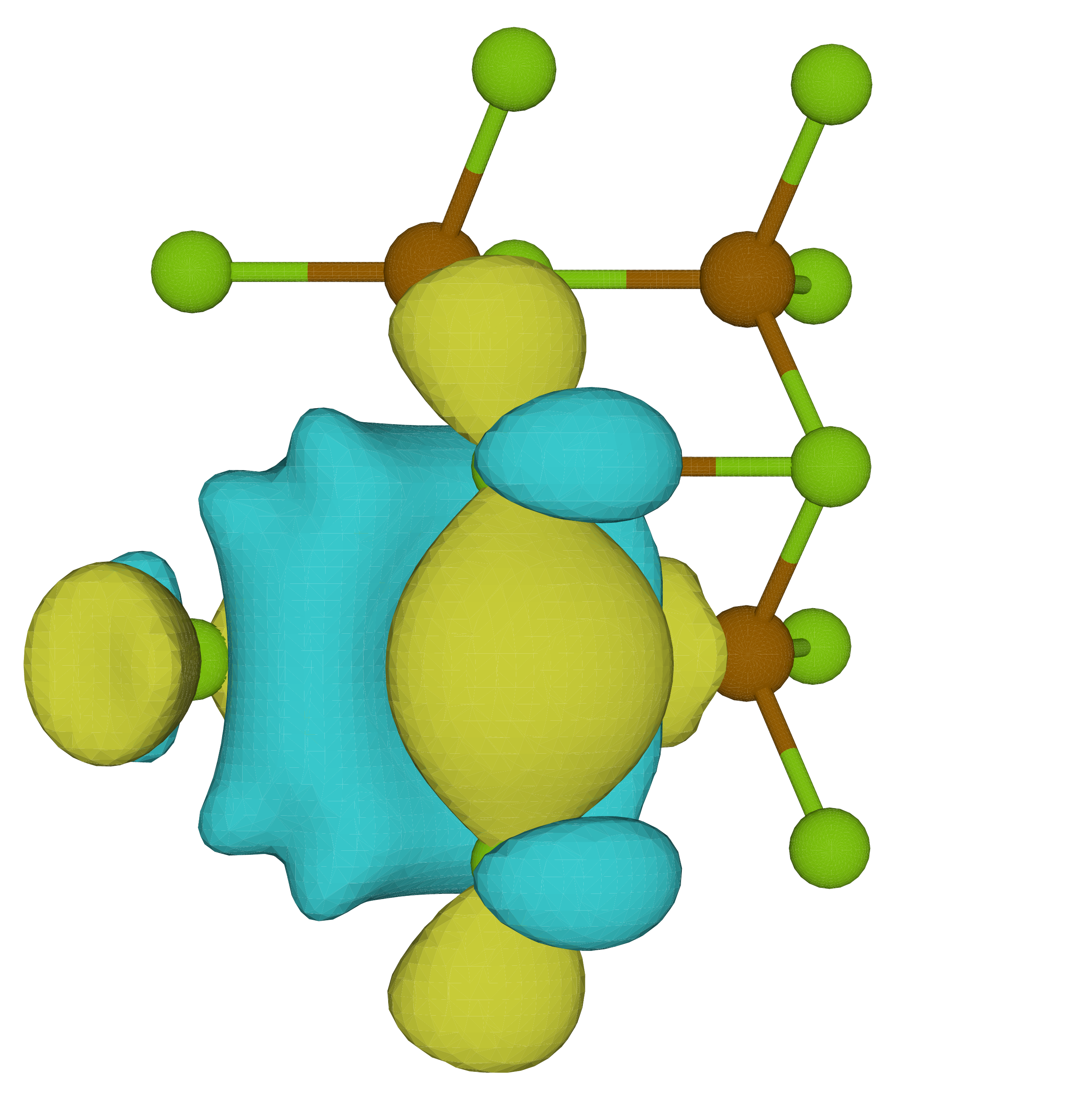}}
\subfloat[]{\includegraphics[width=7cm]{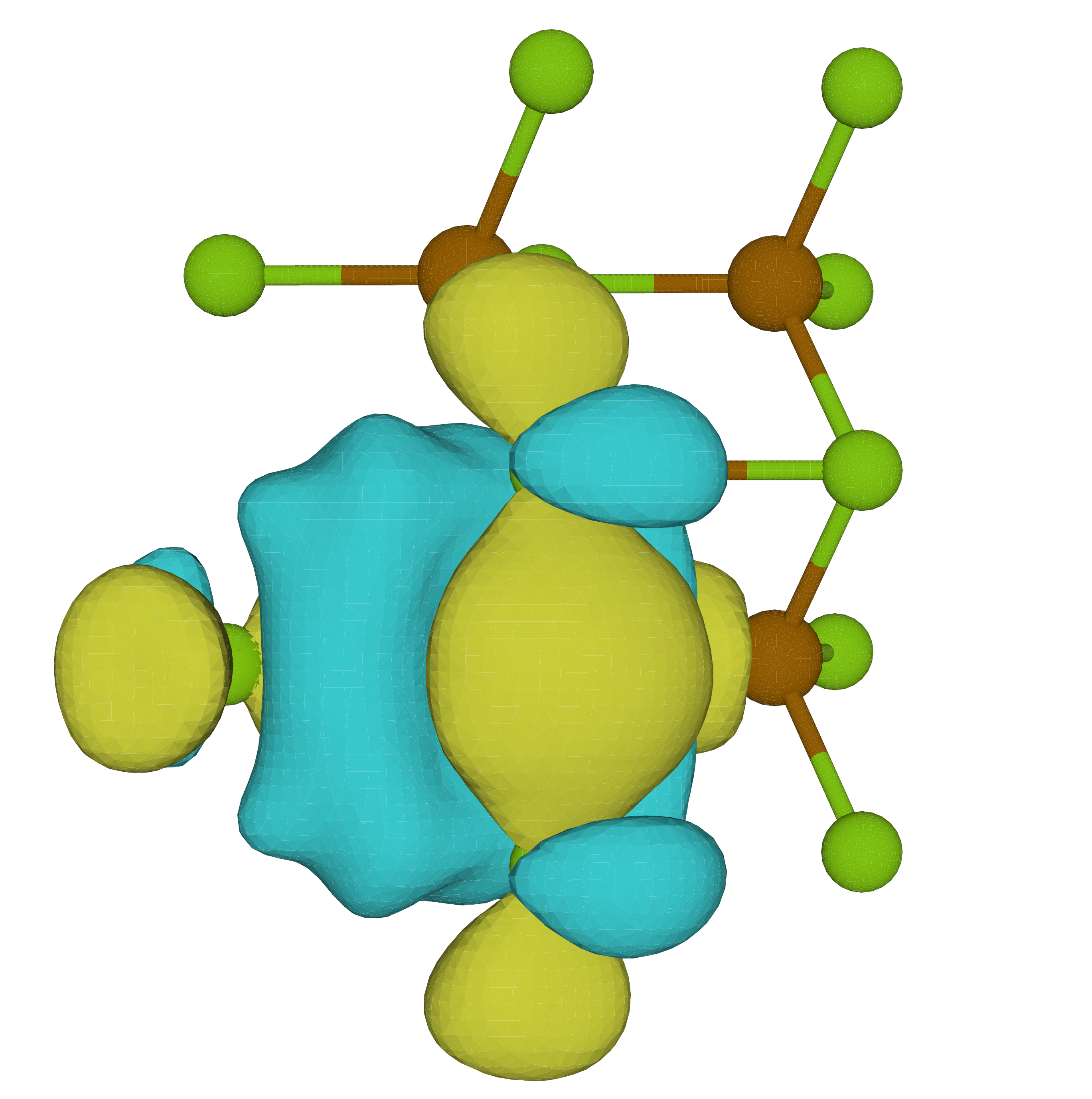}} 
\caption{\label{fig:WFFeSe} Wannier function of single-layer FeSe generated with VASP and Wannier90 (left), and its compression into Gaussian orbitals (right). Positive and negative iso-surfaces corresponding to $12\%$ of the maximum value are plotted.}
\end{figure}

The Wannier function considered here corresponds to a d$-$type orbital on an Fe atom centered at point $A$ and is transformed according to the one-dimensional A$_1$ representation of $\underline{C}_{2v}$, whose character is given in Table~\ref{tab:characters_C2V_A1}. 
\begin{table}
\centering
\begin{tabular}{|c|c|c|c|c|c|c|c|}
\hline
$\underline{C}_{2v}$ & E	 		& C$_2$ (z) 	& $\sigma_{\rm v}$(xz) 	& $\sigma_{\rm v}$(yz) 	& linear  			& quadratic  			& cubic  \\	
									& 				& 						& 											& 								&  functions 	& functions 			& functions \\ 
									\hline
A$_1$ 							& +1 		& +1 				& +1 									& +1 						&  $z$ 			& $x^2$, $y^2$, $z^2$ & $z^3$, $x^2z$, $y^2z$ \\
\hline
\end{tabular}
\caption{Character of the A$_1$ representation of the group $\underline{C}_{2v}$. \label{tab:characters_C2V_A1}}
\end{table}
Graphical representations of the original Wannier function and of its compression into Gaussian orbitals are given in Figure~\ref{fig:WFFeSe}. The decays of the $L^2$ and $H^1$-norms of the residual along the iterations of our implementation of the orthogonal greedy algorithm minimizing the $H^1$-norm of the residual are plotted in Figure~\ref{fig:FeSe_Si_decay}.

\subsection{Diamond-phase silicon}
\begin{figure}[!htb]
\centering
\includegraphics[width=0.5\linewidth]{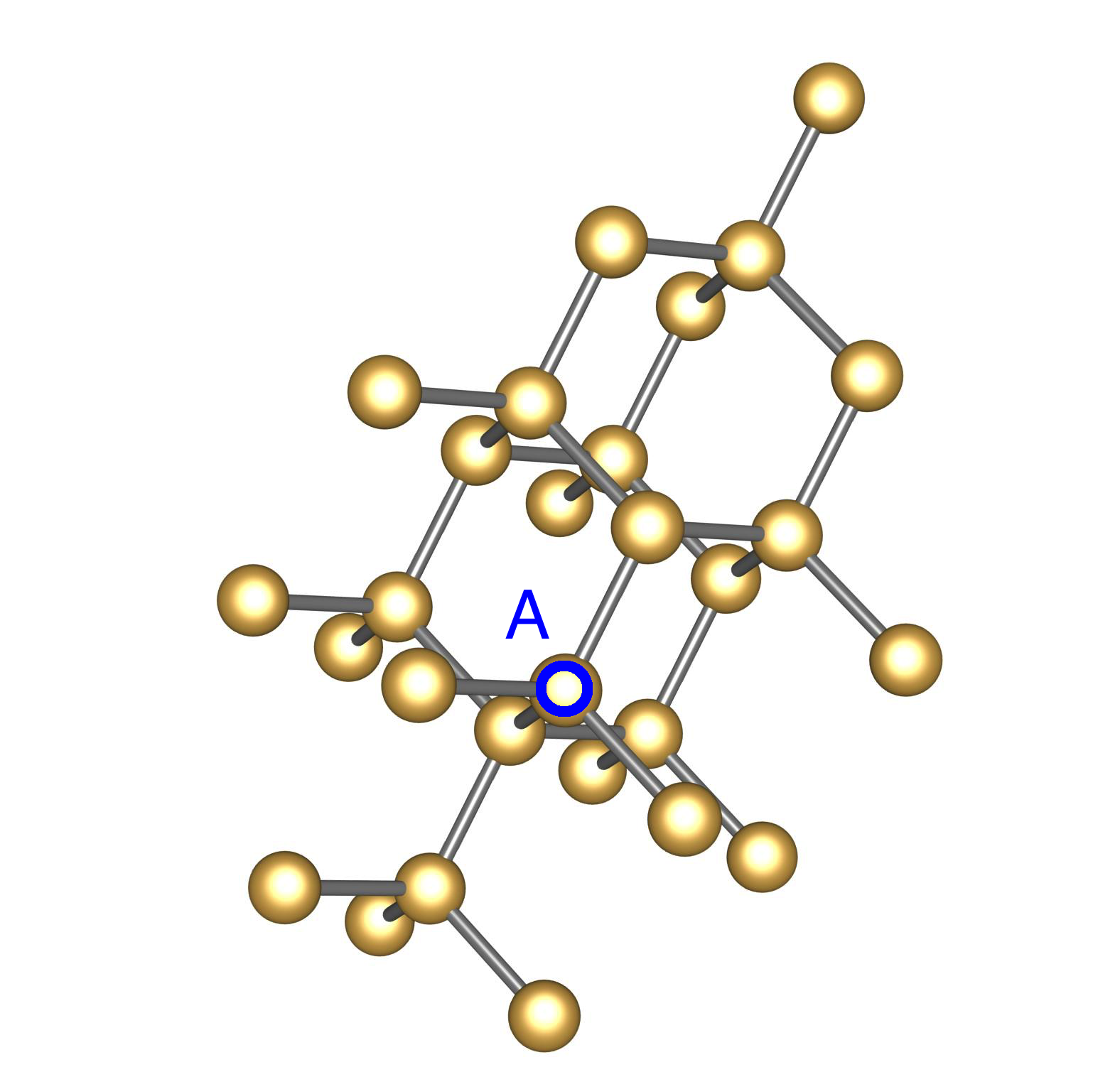} 
\caption{\label{fig:Si_sym} Crystalline structure of Silicon. The brown balls represent Si atoms and the spotted point $A$ corresponds to the high-symmetry point where the Wannier function is centered. }
\end{figure}

The space group of diamond-phase silicon is
\begin{equation*}
G = \mbox{Fd3m} := \cR \rtimes \underline{O}_{h}
\end{equation*}
where $\cR$ is the Bravais lattice of $\R^3$ defined as
\begin{equation}\label{eq:Bravais_graphene}
\cR = \Z a \left( \begin{array}{c} 1 \\0 \\1  \end{array} \right)+ \Z a \left( \begin{array}{c} 1 \\1 \\0 \end{array}  \right) + \Z a \left( \begin{array}{c} 0 \\1 \\1 \end{array} \right), 
\end{equation}
where $a > 0$ is the lattice parameter. The group $\underline{O}_h$ is of order $48$ and has $10$ irreducible representations. The Wannier function considered here corresponds a p$_y-$type orbital centered at the high-symmetry point $A$ represented in Figure~\ref{fig:Si_sym} whose symmetry group is $G_A = \underline{C}_{2v}$.

\begin{figure}[!htb]
\centering
\subfloat[]{\includegraphics[width=7cm]{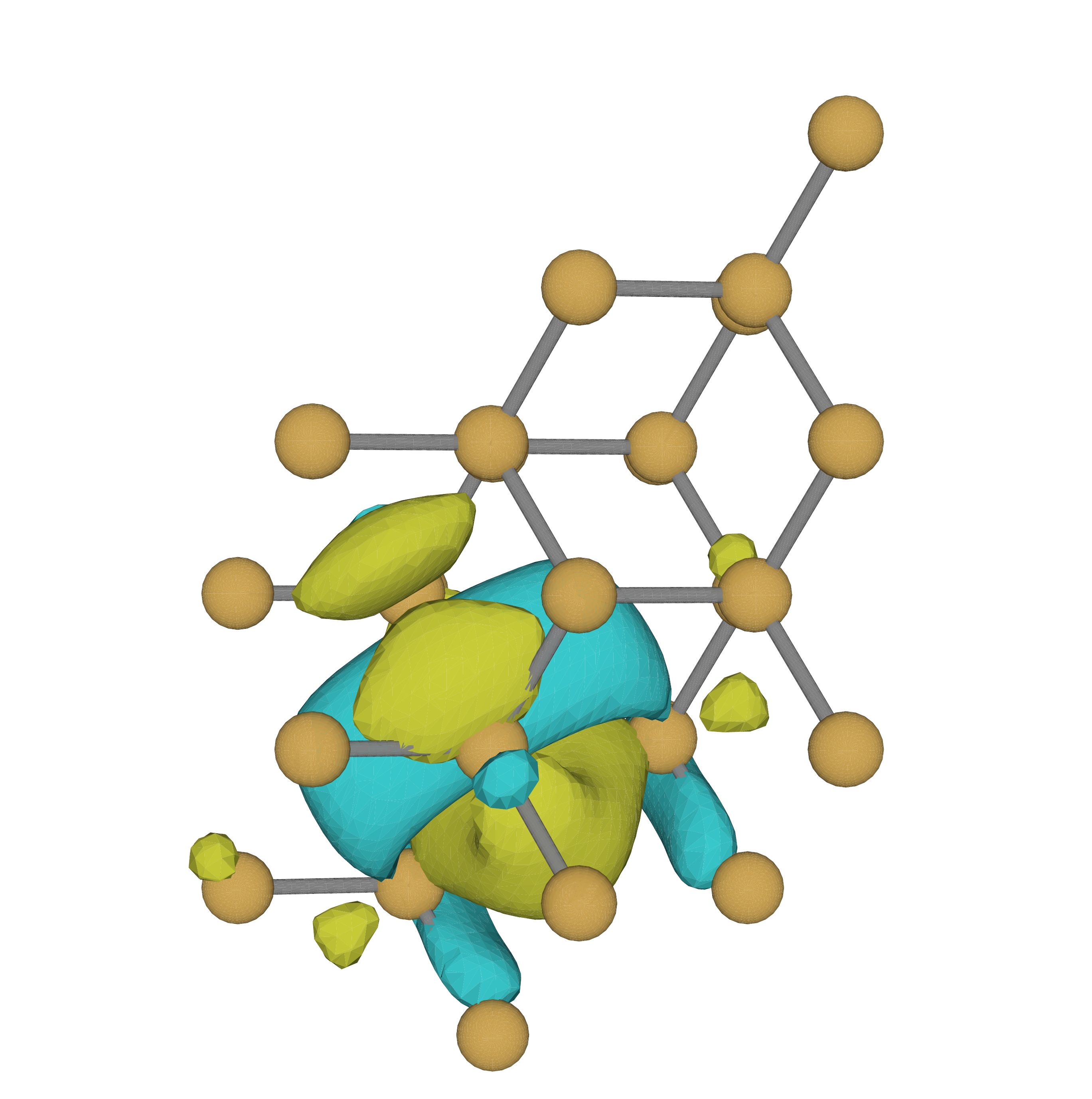}} 
\subfloat[]{\includegraphics[width=7cm]{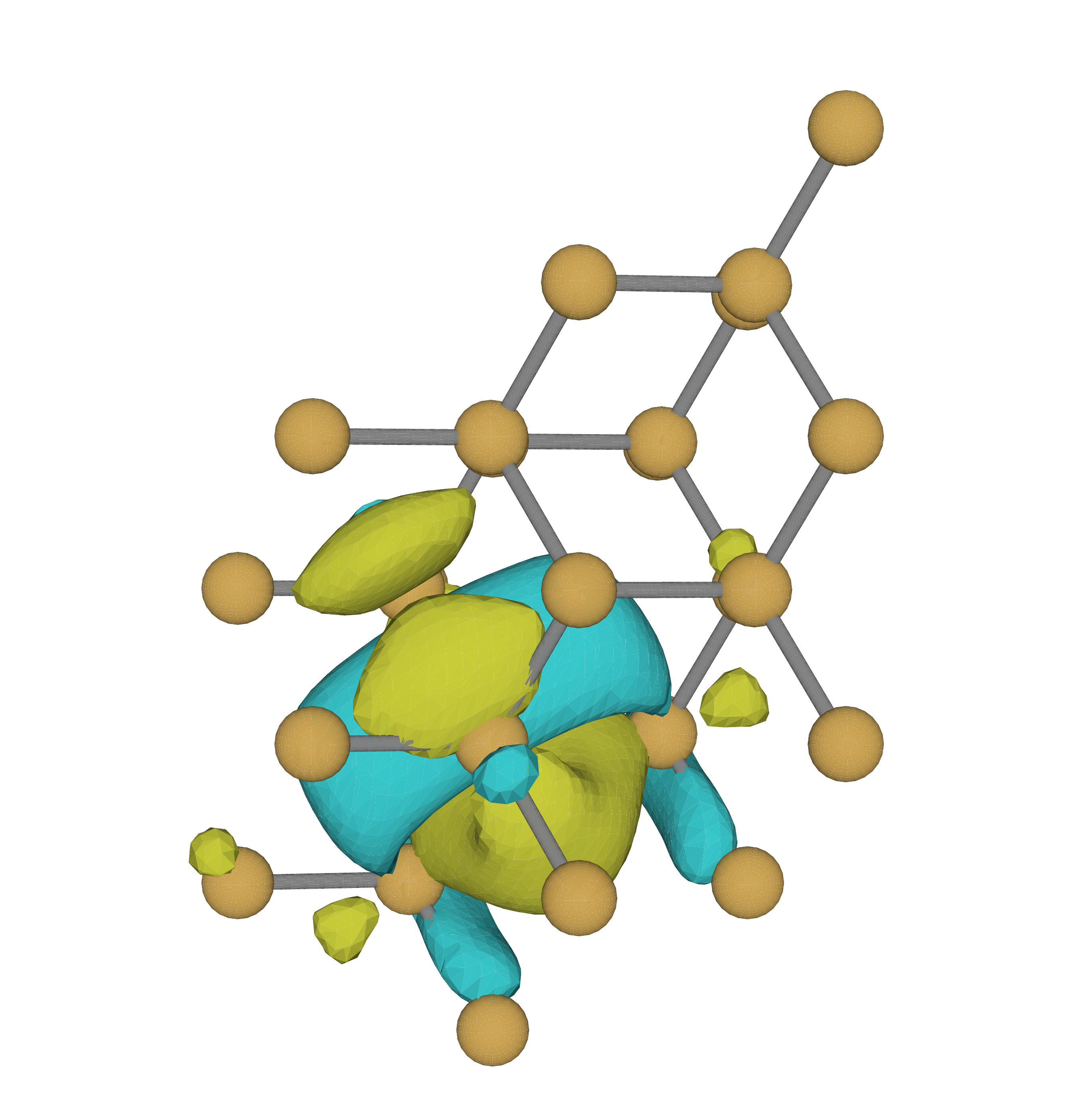}} 
\caption{\label{fig:WFSi} Wannier function of bulk Silicon (diamond phase) generated by Wannier90 (left), and its compression into Gaussian orbitals (right). Positive and negative iso-surfaces corresponding to $10\%$ of the maximum value are plotted.}
\end{figure}

\begin{figure}[!htb]
\centering
\subfloat{\includegraphics[width=0.5\linewidth]{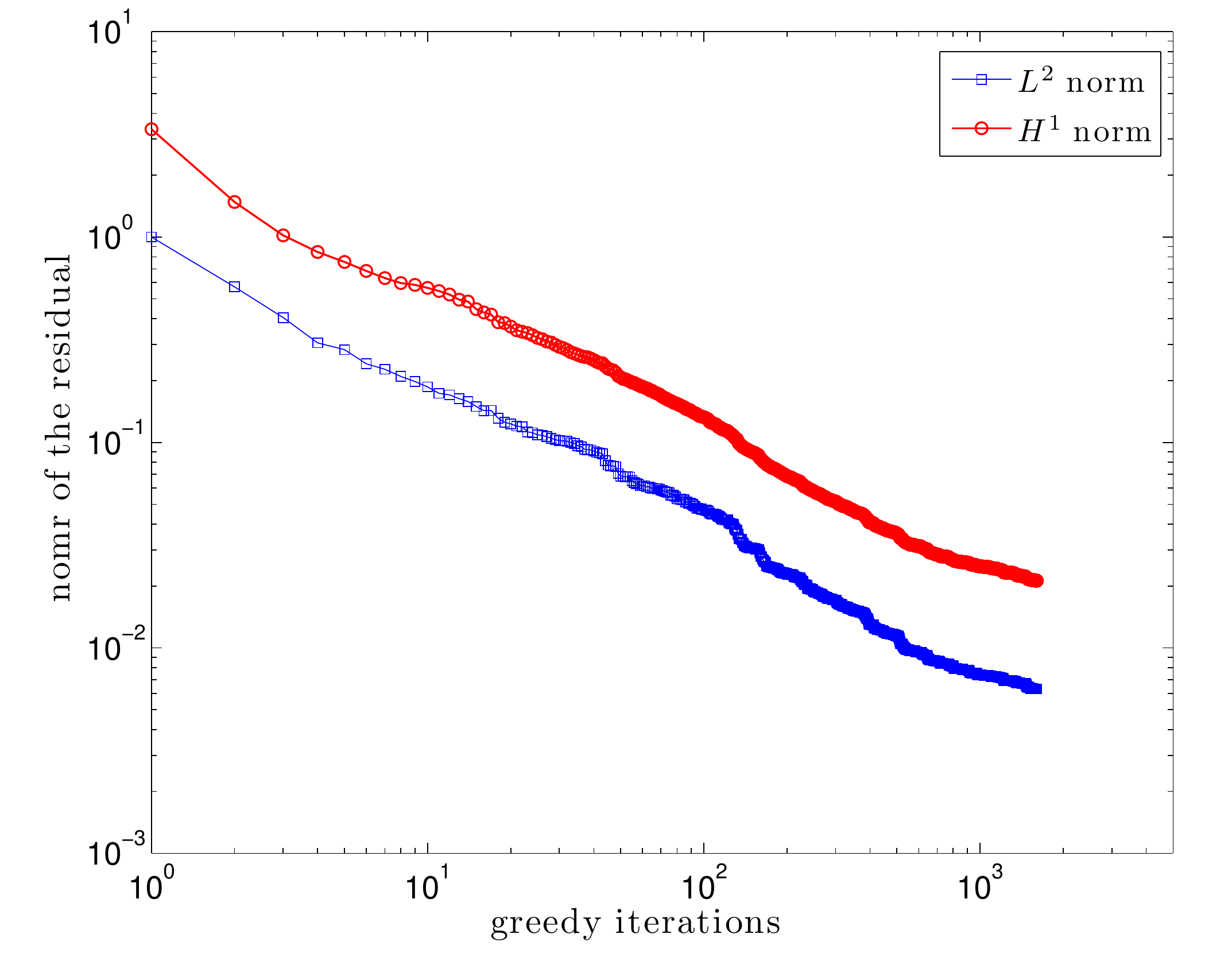}} 
\subfloat{\includegraphics[width=0.5\linewidth]{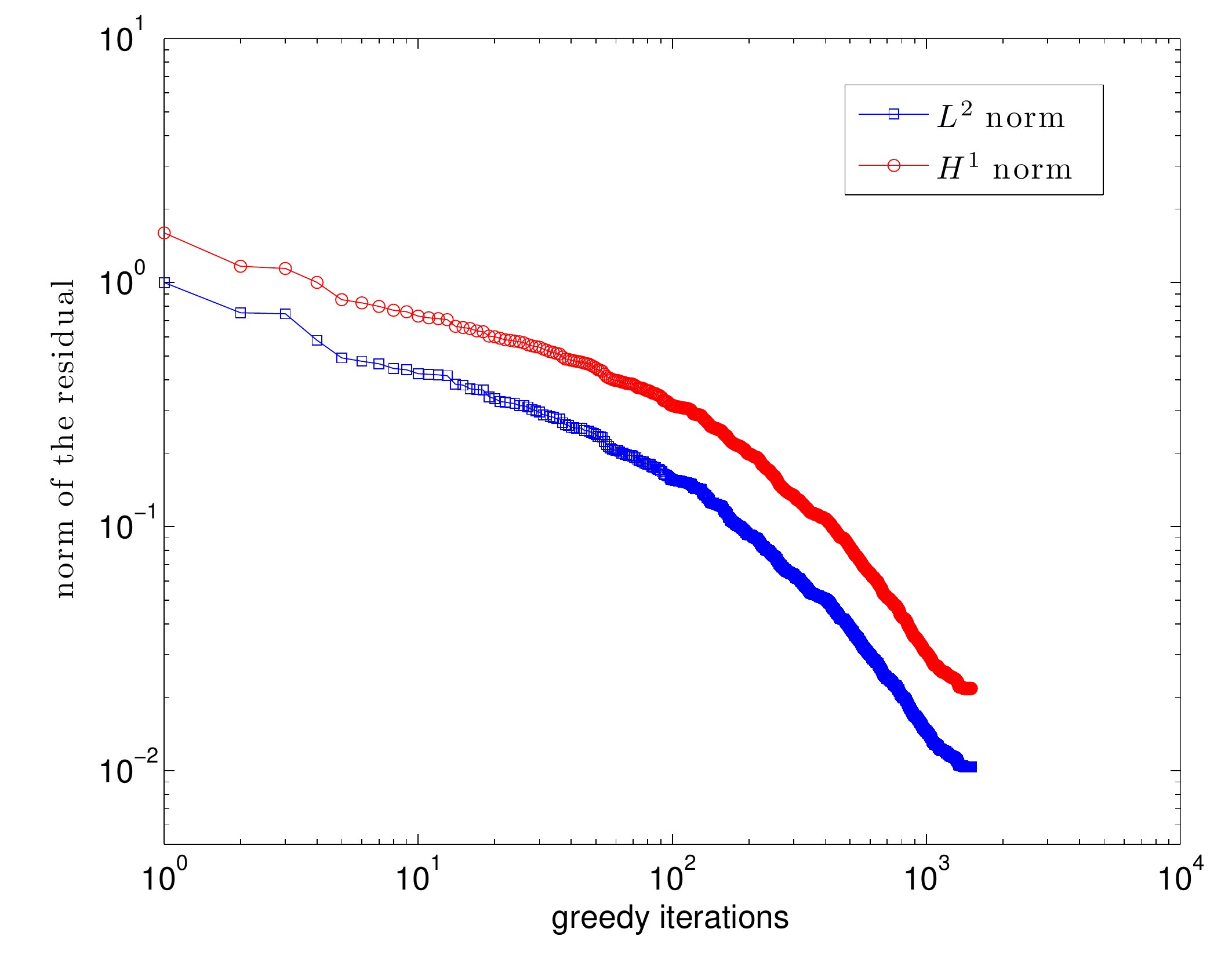}} 
\caption{\label{fig:Graphene_error} Decays of the $L^2$ and $H^1$-norms of the residual for our implementation of the orthogonal greedy algorithm minimizing the $H^1$-norm of the residual (left: FeSe, right: diamond-phase sillicon)\label{fig:FeSe_Si_decay}}
\end{figure}

It is transformed according to the one-dimensional irreducible representation A$_1$ of the  group $\underline{C}_{2v}$.  Let us mention the following point : since the basis $\hat{\textbf{x}} = (1,0,1)$, $\hat{\textbf{y}} = (1,1,0)$ and $\hat{\textbf{z}} (0,1,1)$ is not orthonormal in $\R^3$, the symmetry operators $C_2(z)$, $\sigma_v(xz)$ and $\sigma_v(yz)$ must be adapted to this geometry. Indeed, the two-fold rotation $C_2$ is about the axis of direction $(0,1,1)$ and the two reflexions $\sigma_v$ are defined with respect to the planes $\cP_1$ and $\cP_2$ of cartesian equations $x + z = 0$ and $y+z = 0$ respectively.  
Graphical representations of the original Wannier function and of its compression into Gaussian orbitals are given in Figure~\ref{fig:WFSi}. The decays of the $L^2$ and $H^1$-norms of the residual along the iterations of our implementation of the greedy algorithm aiming at constructing $H^1$-norm approximations of the Wannier function are plotted in Figure~\ref{fig:FeSe_Si_decay}.

%% Tim's suggestion 
%\begin{figure}[!htb]
%\centering
%\begin{tabular}{cc}
%\includegraphics[width=0.4\linewidth]{FeSe_crystal_top.png}  \quad & \quad
%\includegraphics[width=0.4\linewidth]{Si_crystal.png} \\
%\includegraphics[width=0.4\linewidth]{FeSe_true_white.eps} \quad & \quad
%\includegraphics[width=0.4\linewidth]{Si_true_white.eps} \\
%\includegraphics[width=0.4\linewidth]{FeSe_compress_white.eps} \quad & \quad
%\includegraphics[width=0.4\linewidth]{Si_compress_white.eps} \\
%\end{tabular}
%\caption{Tim's suggestion ! FeSe and Si }
%\end{figure}

\section*{Acknowledgments}
This work was supported in part by ARO MURI Award W911NF-14-1-0247.  

\newpage 
\newpage 
\newpage 
\section*{Appendix: symmetry-adapted Wannier functions}
\subsection*{A.1 $\;$ Space group of a periodic material}

Consider a periodic material with $M$ nuclei of charges $z_1, \cdots, z_M$ per unit cell. The nuclear charge distribution in the material is of the form
$$
\nu = \sum_{\bR \in \cR} \sum_{m=1}^M z_m \delta_{\bR_m+\bR},
$$
where $\cR$ is the Bravais lattice of the crystal (embedded in $\R^3$ if the material is a 2D material), $\delta_\ba$ the Dirac mass at point $\ba \in \R^3$, and $\bR_1, \cdots, \bR_M \in \R^3$ the positions of the nuclei laying in the unit cell. The space group $G = \cR \rtimes G_{\rm p}$ of the crystal is the semidirect product of $\cR$ and a finite point group $G_{\rm p}$ (a finite subgroup of $O(3)$). Recall that the composition law in $\cR \rtimes G_{\rm p}$ is defined as 
$$
\forall g_1=(\bR_1,\Theta_1), \quad g_2=(\bR_2,\Theta_2), \quad g_1g_2=(\Theta_1\bR_2+\bR_1,\Theta_1\Theta_2),
$$
and that the natural representation of $G$ in $\R^3$ is given by
$$
\forall g=(\bR,\Theta) \in G, \quad \forall \br \in \R^3, \quad \hat g \br =  \widehat{(\bR,\Theta)}\br = \Theta\br+\bR.
$$
Note that 
$$
\forall g=(\bR,\Theta) \in G, \quad g^{-1} = (-\Theta^{-1}\bR, \Theta^{-1}) \quad \mbox{and} \quad \forall \br \in \R^3, \quad\hat g^{-1}\br =  \Theta^{-1}(\br-\bR). 
$$
The space group of the crystal is the largest group (for an optimal choice of the origin of the Cartesian frame) leaving $\nu$ invariant:
$$
\forall g \in G, \quad \widehat g \nu := \sum_{\bR \in \cR} \sum_{m=1}^M z_m \delta_{\widehat g(\bR_m+\bR)} = \nu.
$$

The group $G$ has a natural unitary representation $\Pi=(\Pi_g)_{g \in G}$ on $L^2(\R^3)$ defined by 
$$
\forall g =(\bR,\Theta) \in G, \quad  (\Pi_g \psi)(\br) = \psi(\hat g^{-1} \br) = \psi( \Theta^{-1}(\br-\bR)).
$$
Denoting by $E$ the identity matrix of rank $3$, and by $\tau=(\tau_\ba)_{\ba \in \R^3}$ the natural unitary representation on $\R^3$ on $L^2(\R^3)$ defined by
$$
\forall \ba \in \R^3, \quad \forall \phi \in L^2(\R^3), \quad (\tau_\ba \phi)(\br) = \phi(\br-\ba),
$$
we have $\Pi_{(\bR,E)} = \tau_\bR$ for all $\bR \in \cR$, so that $(\tau_\bR)_{\bR \in \cR}$ is an abelian subgroup of $\Pi$.

\subsection*{A.2 $\;$ Bloch transform}

Let us now recall the basics of Bloch theory. We denote by $\Gamma$ a unit cell of the Bravais lattice $\cR$, by 
$$
L^2_{\rm per}(\Gamma):= \left\{ u \in L^2_{\rm loc}(\R^3,\C), \; u \; \cR\mbox{-periodic} \right\}, \quad \langle u | v \rangle_{L^2_{\rm per}}:= \int_\Gamma \overline{u(\br)} \, v(\br) \, d\br,
$$
the Hilbert space of locally square-integrable $\cR$-periodic functions $\C$-valued functions on $\R^3$, by $\cR^\ast$ the dual lattice of $\cR$ and by $\Gamma^\ast$ the first Brillouin zone. The Bloch transform associated with $\cR$ (see {\it e.g.} \cite[Section~XIII.16]{Reed_Simon_4}) is the unitary transform
$$
L^2(\R^3,\C) \ni \phi \mapsto (\phi_\bk)_{\bk \in \Gamma^\ast} \in \cH = \fint_{\Gamma^\ast}^\oplus L^2_{\rm per}(\Gamma) \, d\bk
$$
where $\fint_{\Gamma^\ast}$ is a notation for the normalized integral $|\Gamma^\ast|^{-1} \int_{\Gamma^\ast}$, where $\cH$ is endowed with the inner product
$$
\langle (\phi_\bk)_{\bk \in \Gamma^\ast} | (\psi_\bk)_{\bk \in \Gamma^\ast}  \rangle_\cH = \fint_{\Gamma^\ast}  \langle \phi_\bk | \psi_\bk \rangle_{L^2_{\rm per}} \, d\bk,
$$
and where, for a smooth fast decaying function $\phi$, the periodic function $\phi_\bk$ is given by
$$
\phi_\bk(\br) = \sum_{\bR \in \cR} \phi(\br+\bR) e^{-i\bk\cdot(\br+\bR)}.
$$
The original function $\phi$ is recovered from its Bloch transform using the inversion formula
$$
\phi(\br) = \fint_{\Gamma^\ast} \phi_\bk(\br) \, e^{i \bk \cdot \br} \, d\bk.
$$

Consider a one-body Hamiltonian 
$$
H = - \frac 12 \Delta + V_{\rm per}, \quad V_{\rm per} \in L^2_{\rm per}(\Gamma),
$$
describing the electronic properties of the material (we ignore spin for simplicity). In the absence of symmetry breaking, $H$ commutes with all the unitary operators in $\Pi=(\Pi_g)_{g \in G}$. In particular, $H$ commutes with the translations $\tau_\bR$, $\bR \in \cR$, and is therefore decomposed by the Bloch transform:
$$
H = \fint_{\Gamma^\ast} H_\bk \, d\bk,
$$
meaning that there exists a family $(H_\bk)_{\bk \in \Gamma^\ast}$ of self-adjoint operators on $L^2_{\rm per}(\Gamma)$ such that for any $\phi$ in the domain of $H$, $\phi_\bk$ is almost everywhere in the domain of $H_\bk$ and 
$$
(H\phi)_\bk = H_\bk \phi_\bk.
$$
It is well-known that
$$
H_\bk = \frac 12 \left( -i \nabla + \bk \right)^2 + V_{\rm per} =  - \frac 12 \Delta - i \bk \cdot \nabla + \frac 12 |\bk|^2 + V_{\rm per}.
$$
The operator $H_\bk$ can in fact be defined for any $\bk \in \R^3$, and it holds
\begin{equation}\label{eq:covariance_1}
\forall \bk \in \R^3, \quad \forall \bK \in \cR^\ast, \quad H_{\bk+\bK} = V_\bK H_\bk V_\bK^\ast,
\end{equation}
where $V_\bK$ is the unitary operator on $L^2_{\rm per}(\Gamma)$ defined by
$$
\forall u \in L^2_{\rm per}(\Gamma), \quad (V_\bK u)(\br) = e^{-i \bK \cdot \br} u(\br).
$$
As a consequence, for all $\bk \in \R^3$ and $\bK \in \cR^\ast$, $H_\bk$ and $H_{\bk+\bK}$ are unitary equivalent, and therefore have the same spectrum.
Not every $\Pi_g$ {\it a priori} commutes with the translation operators $\tau_\bR$, $\bR \in \cR$. The operator $\Pi_g$ is therefore not in general decomposed by the Bloch transform. On the other hand, denoting by $U=(U_\Theta)_{\Theta \in G_p}$ the natural unitary representation of $G_p$ in $L^2_{\rm per}(\Gamma)$ defined by 
$$
\forall \Theta \in G_{\rm p}, \quad \forall u \in L^2_{\rm per}(\Gamma), \quad (U_\Theta u)(\br) = u(\Theta^{-1}\br),
$$
the Bloch representation of the operator $\Pi_g$, $g=(\bR,\Theta) \in G$, has a simple form:
$$
[\Pi_{(\bR,\Theta)}]_{\bk,\bk'} = e^{-i \bk \cdot \cR} U_\Theta \delta_{\bk',\Theta^{-1}\bk},
$$
that is:
$$
[\Pi_{(\bR,\Theta)}\phi]_\bk (\br) = e^{-i \bk \cdot \cR} \phi_{\Theta^{-1}\bk}(\Theta^{-1}\br).
$$
Since $H$ commutes with all the $\Pi_g$'s, this implies that the family $(H_\bk)_{\bk \in \Gamma^\ast}$ satisfies the covariance relation
$$
\forall \bk \in \R^d, \quad \forall \Theta \in G_p, \quad H_{\Theta\bk}=U_\Theta H_\bk U_\Theta^\ast.
$$
For each $\bk \in \R^3$, the operator $H_\bk$ is self-adjoint on $L^2_{\rm per}(\Gamma)$ and is bounded below. If $\cR$ is a three-dimensional lattice (3D crystal), then $H_\bk$ has a compact resolvent and its spectrum is purely discrete. If $\cR$ is a two-dimensional lattice (2D material), then the essential spectrum of $H_\bk$ is a half-line $[\Sigma_\bk,+\infty)$.

\subsection*{A.3 $\;$ Symmetry-adapted Wannier functions}

We assume here that $H$ has a finite number $n \ge 1$ of bands isolated from the rest of the spectrum, that is that there exist two continuous $\R$-valued $\cR$-periodic functions $\bk \mapsto \mu_-(\bk)$ and $\bk \mapsto \mu_+(\bk)$ such that $\mu_-(\bk) < \mu_+(\bk)$, $\mu_\pm(\bk) \notin \sigma(H_\bk)$ and $\tr(\1_{[\mu_-(\bk),\mu_+(\bk)]}(H_\bk))=n$ for all $\bk \in \R^3$. We denote by $\epsilon_{1,\bk} \le \epsilon_{2,\bk} \le \cdots \le \epsilon_{n,\bk}$ the eigenvalues of $H_\bk$ laying in the range $[\mu_-(\bk),\mu_+(\bk)]$ (counting multiplicities). The functions $\bk \mapsto \epsilon_{n,\bk}$ are Lipschitz continuous, and, in view \eqref{eq:covariance_1}, are also $\cR$-periodic.

\medskip

A generalized Wannier function associated to these $n$ bands is a function of the form
$$
\forall \br \in \R^3, \quad W(\br) = \fint_{\Gamma^\ast} u_\bk(\br) \, e^{i \bk \cdot \br} \, d\bk, \quad u_\bk \in \mbox{Ran}(\1_{[\mu_-(\bk),\mu_+(\bk)]}(H)), \quad \|u_\bk\|_{L^2_{\rm per}}=1.
$$

\medskip

Let $\bq$ be a site of the unit cell of the crystalline lattice\footnote{Here the lattice is not in general a Bravais lattice. For graphene and hBN, this is a honeycomb lattice.}. We denote by 
$$
G_\bq=\left\{ g=(\bR,\Theta) \in G \; | \; \hat g \bq=\Theta\bq+\bR = \bq \right\}
$$
the finite subgroup of $G$ leaving $\bq$ invariant. The point $\bq$ is called a high-symmetry point if $G_\bq$ is not trivial. Setting $\bR_\Theta = \bq - \Theta \bq$, we have
$$
G_\bq=\left\{ g=(\bR_\Theta,\Theta), \; \Theta \in G_\bq^0 \right\},   
$$
where $G_\bq^0$ is a subgroup of $G_{\rm p}$.

\medskip

A symmetry-adapted Wannier function centered at a high-symmetry point $\bq$ is a Wannier function $W$ such that 
\begin{enumerate}
\item the finite-dimensional space
$$
\cH_{W,\bq}:=\mbox{Span} \left( \Pi_g W, \; g \in G_\bq \right)
$$
is $\Pi_g$-invariant for any $g \in G_\bq$;
\item $(\Pi_g|_{\cH_{W,\bq}})_{g \in G_\bq}$ defines an irreducible unitary representation $\beta$ of $G_\bq$. 
\end{enumerate}
Let $n_\beta:=\mbox{dim}(\cH_{W,\bq})$ be the dimension of this representation and $(W^{(\beta)}_{i,1})_{1 \le i \le n_\beta}$ be a basis of $\cH_{W,\bq}$ such that $W^{(\beta)}_{1,1}=W$. Let $(d^{\beta}(\Theta))_{\Theta \in G_\bq^0} \in (\C^{n_\beta \times n_\beta})^{n_\bq}$ be the matrix representation of the group $G_\bq^0$ in 
$$
\cH_{W,\bq}^0:=\mbox{Span} \left( \Pi_\Theta \tau_{-\bq} W, \; \Theta \in G_\bq^0 \right), \quad \mbox{where} \quad \Pi_\Theta:=\Pi_{({\bold 0},\Theta)}.
$$
We therefore have
$$
\forall \Theta \in G_\bq^0, \quad \Pi_\Theta \left( \tau_{-\bq} W^{(\beta)}_{i,1} \right) = \sum_{i'=1}^{n_\beta} d_{i',i}^{(\beta)}(\Theta)  \left( \tau_{-\bq}  W^{(\beta)}_{i',1} \right),
$$
so that
$$
\forall (\bR_\Theta,\Theta) \in G_\bq, \quad \Pi_{(\bR_\Theta,\Theta)}  W^{(\beta)}_{i,1} = \sum_{i'=1}^{n_\beta} d_{i',i}^{(\beta)}(\Theta)   W^{(\beta)}_{i',1} .
$$
If the representation $\beta$  is one-dimensional ($n_\beta=1$), then $(d^{\beta}(\Theta))_{\Theta \in G_\bq^0}$ is the character of the corresponding representation of $G_\bq^0 \subset G_p$ in $\cH_{W,\bq}^0$.

\medskip

Let $J=|G_{\rm p}|/|G_\bq| \in \N^\ast$. Then, there exist $(g_j)_{1 \le j \le J} \in G^J$ such that 
$$
G = \sum_{j=1}^J \sum_{\bR \in \cR} (\bR|E)g_j G_\bq.
$$
More precisely, there exist $(g_j)_{1 \le j \le J} \in G^J$ such that 
\begin{itemize}
\item for each $1 \le j \le J$, $\bq_j:=\hat g_j \bq \in \Gamma$;
\item any $g \in G$ can be decomposed in a unique way as
$$
g = (\bR|E)g_jg_\bq
$$
for a unique triplet $(\bR,j,g_\bq) \in \cR \times |[1,J]| \times G_\bq$. 
\end{itemize}

\medskip

\noindent
For each $1 \le i \le n_\beta$, $1 \le j \le J$ and $\bR \in T$, we set
$$
W_{i,j,\bR}^{(\beta)} = \Pi_{(\bR|E)g_j} W_{i,1}^{(\beta)} ,
$$
and we then define 
$$
\cH_W= \overline{\mbox{Span} \left( W_{i,j,\bR}^{(\beta)}, \; 1 \le i \le n_\beta, \; 1 \le j \le J, \; \bR \in \cR \right\}}.
$$
In other words, $\cH_W$ is the closure of the vector space generated by the mother SAWF $W$ and all the SAWFs obtained by letting the elements of $G$ act on $W$.

The space $\cH_W \subset H^2(\R^3)$ is both $H$-invariant and $\Pi$-invariant, and for any $g \in G$, the action of $\Pi_g$ on $W_{i,j,\bR}^{(\beta)}$ can be computed as follows. Let $(\bR',j',g'_\bq)$ the unique element of $\cR \times |[1,J]| \times G_\bq$ such that $g (\bR|E)g_j =  (\bR'|E)g_{j'}g'_{\bq}$. We have
\begin{align*}
\Pi_g W_{i,j,\bR}^{(\beta)} &=   \Pi_g  \Pi_{(\bR|E)g_j} W_{i,1}^{(\beta)} = \Pi_{g (\bR|E)g_j} W_{i,1}^{(\beta)} = \Pi_{(\bR'|E)g_{j'}g'_{\bq}} W_{i,1}^{(\beta)} \\
&= \Pi_{(\bR'|E)g_{j'}} \Pi_{g'_{\bq}} W_{i,1}^{(\beta)} = \Pi_{(\bR'|E)g_{j'}} \left( \sum_{i'=1}^{n_\beta} d_{i',i}^{(\beta)}(\Theta'_{q})   W^{(\beta)}_{i',1} \right) \\
&= \sum_{i'=1}^{n_\beta} d_{i',i}^{(\beta)}(\Theta'_{q})  W^{(\beta)}_{i',j',\bR'}. 
\end{align*}
The index $j'$ is the unique integer in the range $|[1,J]|$ such that 
$$
\hat g(\bq_j+\bR) \in \bq_{j'}+\cR.
$$
The explicit expressions of $\bR'$ and $\Theta'_{\bq}$ as functions of $(\bR,j)$ and $g=(\bR,\Theta)$ are the following
$$
\Theta'_\bq = \Theta_{j'}^{-1} \Theta \Theta_j, \quad \bR'= \hat g \bq_j- \bq_{j'}+ \Theta \bR.
$$

Constructing a basis of SAWFs for the $n$ bands defined by the functions $\mu_-$ and $\mu_+$ amounts to finding $s \in \N^\ast$ high-symmetry points $\bq_1, \cdots, \bq_s$, and $s$ SAWFs Wannier functions $W_1, \cdots, W_s$ respectively centered at the points $\bq_1, \cdots, \bq_s$, such that
$$
\fint_{\Gamma^\ast}^\oplus \mbox{Ran}\left(\1_{[\mu_-(\bk),\mu_+(\bk)]}(H) \right) \, d\bk = \cH_{W_1} \oplus \cdots \oplus \cH_{W_s}.
$$
This is the purpose of the numerical method introduced in~\cite{S13}.

 \bibliographystyle{plain}
 \bibliography{wanbib}
\end{document}